\documentclass[useAMS,usenatbib,usegraphicx]{mn2e}
\usepackage{ gensymb }

\usepackage[fleqn]{amsmath}

\newcommand{\dd}{{\rm d}}
\newcommand{\Pm}{\mathcal{P}}
\newcommand\bnabla{{\bmath\nabla}}
\newcommand\bB{{\bmath B}}
\newcommand\be{{\bmath e}}
\newcommand\bv{{\bmath v}}
\newcommand\bfI{\mathbf{I}}
\newcommand\bfT{\mathbf{T}}

\newcommand\half{{\textstyle\frac{1}{2}}}

\newcommand\rmd{\mathrm{d}}
\newcommand\rme{\mathrm{e}}

\newcommand\rms{\mathrm{s}}

\newcommand\rmT{\mathrm{T}}
\newcommand\f{\frac}
\newcommand\p{\partial}
\newcommand\cst{\mathrm{constant}}

\usepackage{epsfig}
\usepackage{amsmath}

\title[Magnetic flux transport in accretion discs]
{Transport of magnetic flux and the vertical structure of accretion discs: I. Uniform diffusion coefficients}

\author[J\'er\^ome Guilet and Gordon I. Ogilvie]
{J\'er\^ome Guilet and Gordon I. Ogilvie\\
Department of Applied Mathematics and Theoretical Physics,
University of Cambridge, Centre for Mathematical Sciences,\\
Wilberforce Road, Cambridge CB3 0WA
}

\begin{document}

\maketitle

\label{firstpage}

\begin{abstract}
Standard models of accretion discs study the transport of mass on a viscous timescale but do not consider the transport of magnetic flux. The evolution of a large-scale poloidal magnetic field is however an important problem because of its role in the launching of jets and winds and in determining the intensity of turbulence. As a consequence, the transport of poloidal magnetic flux should be considered on an equal basis to the transport of mass. In this paper, we develop a formalism to study such a transport of mass and magnetic flux in a thin accretion disc. The governing equations are derived by performing an asymptotic expansion in the limit of a thin disc, in the regime
  where the magnetic field is dominated by its vertical component.
  Turbulent viscosity and resistivity are included, with an arbitrary
  vertical profile that can be adjusted to mimic the vertical
  structure of the turbulence. At a given radius and time, the rates
  of transport of mass and magnetic flux are determined by a
  one-dimensional problem in the vertical direction, in which the
  radial gradients of various quantities appear as source terms. We
  solve this problem to obtain the transport rates and the vertical
  structure of the disc. The present paper is then restricted to the idealised case of uniform diffusion coefficients, while a companion paper will study more realistic vertical profiles of these coefficients. We show the advection of weak magnetic fields
  to be significantly faster than the advection of mass, contrary to
  what a crude vertical averaging might suggest.  This results from
  the larger radial velocities away from the mid-plane, which barely
  affect the mass accretion owing to the low density in these regions
  but do affect the advection of magnetic flux. Possible consequences of this larger accretion velocity include a potentially interesting time-dependence with the magnetic flux distribution evolving faster than the mass distribution. If the disc is not too thin, this fast advection may also partially solve the long-standing problem of too efficient diffusion of an inclined magnetic field. 
  \end{abstract}


\section{Introduction}

The presence of a large-scale magnetic field in an accretion disc has
several potentially important consequences.  It may play an
essential role in the acceleration and collimation of jets and winds
from the disc \citep{blandford82} and in harnessing the rotational
energy of a central black hole \citep{blandford77}.  It also
strongly affects the intensity of magnetohydrodynamic (MHD) turbulence and
the resulting transport of angular momentum within the disc
\citep{balbus98}.

The origin (and existence) of such a large-scale magnetic field is,
however, debated from a theoretical perspective. There are in
principle two possibilities: either the field is created in situ by a
dynamo process, or it is brought in by the gas that is being accreted.
It is still unclear whether a dynamo process can create a significant magnetic field that would be coherent over a
scale comparable to the radius. Indeed the MHD turbulence, presumed to be at the origin of the dynamo, most likely has a radial correlation length of the order of the vertical thickness of the disc, which is usually much smaller the radius  (see the discussion in
\citealt{spruit10}).

In the second possibility, an initially weak field present in the gas
supplied to the disc could be strongly amplified as it is transported
inwards and the magnetic flux accumulates in the central part
of the disc. The transport of magnetic flux is due to two processes: the inward advection by the accretion flow, and the diffusion due to a turbulent resistivity which tends to counteract the advection and therefore transport the magnetic flux outwards. The scenario therefore requires that the field diffuse outwards at a slower rate than it is advected for the total transport to be directed inwards (at least initially). Which of the advection or the diffusion is faster is not obvious a priori, as both
are due to the same phenomenon: the turbulence which is causing an
effective viscosity (enabling advection) and an effective resistivity
(enabling the magnetic field to diffuse). The first theoretical
studies of the evolution of the magnetic field in the presence of
advection and diffusion have found that the diffusion is much faster
than the advection, if the disc is thin and the field lines bend significantly across it \citep{lubow94a,heyvaerts96}. This can be understood by the
following argument. If the angular momentum transport is due to a
turbulent viscosity $\nu$, the advection speed is approximately
\begin{equation}
v_{\rm adv} \sim \f{3}{2}\frac{\nu}{r}.
	\label{eq:vadv}
\end{equation}
The diffusion of the magnetic flux is predominantly due to the bending
of the magnetic field lines across the disc, which is associated with
an azimuthal electric current $J_\phi \sim B_r/(\mu_0H)$, where $H$ is the
scale-height of the disc. The action of the turbulent magnetic diffusivity $\eta$ on this current induces a diffusion speed
 \begin{equation}
v_{\rm diff} \sim \frac{\eta}{H}\frac{B_r}{B_z}.
	\label{eq:vdiff}
\end{equation}
Equating the advection and diffusion speeds, one finds the maximum
inclination of the magnetic field lines that can be induced by the
advection:
\begin{equation}
\frac{B_r}{B_z} \sim \frac{3}{2} \Pm\frac{H}{r},
	\label{eq:vadv_diff}
\end{equation}
where $\Pm \equiv \nu/\eta$ is the turbulent magnetic Prandtl number,
which is usually expected to be of order unity. The bending is thus
very small for a thin disc, unless the magnetic Prandtl number is
unexpectedly large ($\Pm \sim r/H$).

This result is rather disappointing as it is problematic for models of
jets and winds. Indeed, the acceleration of a jet or wind by the
magneto-centrifugal mechanism requires the field lines to bend by more
than $30$ degrees from the vertical direction \citep{blandford82}.
Furthermore, outward bending of the field lines is a natural
consequence of the accumulation of magnetic flux in the central part
of the disc. The result of \citet{lubow94a} thus suggests that the
inward advection of magnetic flux cannot significantly amplify the
magnetic field, and raises doubts on the very existence of a
significant large-scale magnetic
field. 

Various potential solutions to this magnetic diffusion problem have been considered: 
\begin{itemize}
\item A high effective magnetic Prandtl number associated with the
  turbulence could solve the problem. However, theoretical
  expectations \citep{parker71,pouquet76} as well as shearing-box
  simulations of MHD turbulence suggest that the turbulent magnetic
  Prandtl number is of order unity \citep{lesur09,guan09,fromang09}. Note however that the main contribution to the diffusion (the vertical diffusion of a radial field) has not been measured so far. It is also possible that the effects of turbulence are more complicated than is described by effective diffusion coefficients.

\item Breaking of the axial symmetry: \citet{spruit05} proposed a
  scenario allowing a fast advection of the magnetic flux while
  preventing its diffusion. In this scenario a weak magnetic field
  concentrates into strongly magnetised small patches, owing to the
  tendency of turbulent flows to expel magnetic flux (e.g.
  \citealt{weiss66}). The diffusion is slowed down because the strong
  magnetic field prevents turbulence inside the patch. On the other
  hand, rapid advection occurs as the patch loses angular momentum
  through a magneto-centrifugal wind. This interesting idea needs
  however much more work to be confirmed.

\item Vertical structure: the analysis by \cite{lubow94a} uses a
  vertical averaging of the disc, which implicitly assumes that the
  magnetic flux is advected at the same speed as the mass. However,
  \cite{ogilvie01} have questioned the validity of this averaging
  procedure and suggested that the vertical dependence of the radial
  velocity, the density and the resistivity could have an
  influence. One objective of this article is to investigate further this question.
  \citet{bisnovatyi-kogan07,bisnovatyi-kogan12}, \citet{rothstein08}, and
  \citet{lovelace09} also considered the vertical structure of the
  disc, studying the effect of a non-turbulent layer at the surface.
  They suggested that most of the bending of the magnetic field lines
  would take place in the highly conducting non-turbulent layer.  The
  low resistivity of the layer would thus drastically reduce the
  diffusion of the magnetic field. A proper self-consistent calculation of this effect is still missing, however, because the
solutions of \citet{lovelace09} and \citet{bisnovatyi-kogan12} stop at the base of the conducting surface layer. The formalism developed in this article will be used in a companion paper to test this scenario. 
\end{itemize}

The main purpose of this article and the companion one (Guilet \& Ogilvie in preparation, hereafter called paper II) is to clarify how the vertical structure of the disc affects the transport of magnetic flux.  We are
particularly interested in the poloidal magnetic flux threading the
disc, as this quantity satisfies a conservation equation and can be
modified only by advective or diffusive transport, possibly enhanced
by turbulence.  Since the poloidal magnetic field strongly affects the
intensity of turbulence in an accretion disc and is an essential
component in magneto-centrifugal jet launching, the transport of
poloidal flux needs to be considered on an equal basis to the
transport of mass in an accretion disc.

Although future work may involve direct numerical simulations of
turbulent discs, in our present analysis the turbulence is simply
modelled by an effective viscosity and an effective resistivity. The formalism described in Sections~\ref{sec:asymptotic} and Section~\ref{sec:local} allows any vertical profile of these coefficients, which will be used in paper II to study the effect of
the variation of the turbulent intensity with distance from the
mid-plane. The rest of this article however focuses on the idealised case of uniform diffusion coefficients. Under these assumptions and in the limit of a thin accretion
disc, we self-consistently solve for the vertical profile of the mean
velocity and magnetic field. We focus in particular on the regime in
which the magnetic field is dominated by its vertical component. This
is done partly to simplify the analysis and take advantage of a linear
system of equations, and partly in order to avoid the complications of
magneto-centrifugal jet launching from the disc.  A separate paper will
address the launching process \citep{ogilvie12}. 

The plan of this paper is as follows. In Section~\ref{sec:asymptotic},
we use the basic equations of MHD, together with an asymptotic expansion in the limit of a thin
disc, to derive the equations describing the evolution of the mass and
magnetic flux distributions on a viscous or resistive timescale. In
Section~\ref{sec:local}, we focus on the quasi-local problem
describing the vertical structure of the thin disc and enabling the
calculation of the transport rates. This problem is solved for the
special case of uniform diffusion coefficients in
Section~\ref{sec:cstalpha}. The results are discussed and summarised in Section~\ref{sec:discussion}.

\section{An asymptotic expansion within a thin disc}
	\label{sec:asymptotic}

We start from the equations of MHD, in the form
\begin{equation}
  \f{\p\rho}{\p t}+\bnabla\cdot(\rho\bv)=0,
\end{equation}
\begin{eqnarray}
  \lefteqn{\rho\left(\f{\p\bv}{\p t}+\bv\cdot\bnabla\bv\right)=-\rho\bnabla\Phi-\bnabla p}&\nonumber\\
  &&+\f{1}{\mu_0}(\bnabla\times\bB)\times\bB+\bnabla\cdot\bfT,
\end{eqnarray}
\begin{equation}
  \f{\p\bB}{\p t}=\bnabla\times(\bv\times\bB-\eta\bnabla\times\bB),
\end{equation}
\begin{equation}
  \bnabla\cdot\bB=0,
\end{equation}
\begin{equation}
  \bfT=\rho\nu\left[\bnabla\bv+(\bnabla\bv)^\rmT-\f{2}{3}(\bnabla\cdot\bv)\,\bfI\right],
\end{equation}
where $\rho$ is the density, $\bv$ the velocity, $\Phi$ the
gravitational potential, $p$ the pressure, $\bB$ the magnetic field,
$\bfT$ the viscous stress, $\eta$ the magnetic diffusivity, $\nu$ the
kinematic viscosity and $\bfI$ the unit tensor of second rank.  (A
bulk viscosity could be included, but would have no effect on this
problem.)  We neglect self-gravity and do not consider a thermal
energy equation at this stage.

We consider an axisymmetric solution of these equations in cylindrical
polar coordinates $(r,\phi,z)$.  The poloidal part of the magnetic
field can be described in the usual way by a poloidal magnetic flux
function $\psi(r,z,t)$.  Thus
\begin{equation}
  \bB=\bnabla\psi\times\bnabla\phi+B_\phi\,\be_\phi=-\f{1}{r}\be_\phi\times\bnabla\psi+B_\phi\,\be_\phi.
\end{equation}
This representation satisfies the constraint $\bnabla\cdot\bB=0$
automatically.  Poloidal magnetic field lines correspond to curves
$\psi=\cst$ in the $(r,z)$ plane, and (up to an additive constant)
$2\pi\psi$ is the magnetic flux contained within a circular loop at
position $(r,z)$.  The poloidal part of the induction equation can
then be integrated to give
\begin{equation}
  \f{\p\psi}{\p t}+\bv\cdot\bnabla\psi=\eta r^2\bnabla\cdot\left(\f{1}{r^2}\bnabla\psi\right),
\label{dpsidt}
\end{equation}
which shows that the poloidal magnetic flux satisfies a conservation
law and is affected by advection and diffusion.

We are interested in solving these equations both inside and outside a
thin accretion disc that lies close to the plane $z=0$.  In the
absence of magnetic fields, there is a well understood ordering scheme
for thin discs.  If the characteristic angular semithickness of the
disc is $H/r=O(\epsilon)$, where $0<\epsilon\ll1$ is a small
dimensionless parameter, and the Shakura--Sunyaev viscosity parameter
is $\alpha$, then the ratio of the sound speed to the orbital velocity
is $O(\epsilon)$, and the ratio of the radial velocity to the orbital
velocity is $O(\alpha\epsilon^2)$.  The fractional deviation of the
azimuthal velocity from the Keplerian value is $O(\epsilon^2)$, and so
on.  Furthermore, the disc evolves on the viscous timescale, which is
longer than the orbital timescale by $O(\alpha^{-1}\epsilon^{-2})$.
Although this is rarely done, the equations governing thin accretion
discs can be obtained by a formal asymptotic expansion of the basic
equations of fluid dynamics.

For magnetised discs there is more than one possible ordering scheme,
depending on the strength (and orientation) of the magnetic field
\citep{ogilvie97}.  We consider here a situation in which the disc has
comparable values of $\nu$ and $\eta$, with (formally) $\alpha=O(1)$.
The magnetic field is dominated by the vertical component, which has a
magnetic pressure comparable to the gas pressure.  This assumption
allows us to avoid a consideration of magnetocentrifugal jet
launching.

We resolve the internal structure of the thin disc and its slow
evolution in time through the introduction of a rescaled vertical
coordinate
\begin{equation}
  \zeta=\epsilon^{-1}z
\end{equation}
and a time variable
\begin{equation}
  \tau=\epsilon^2t.
\end{equation}
We then propose the following expansion of the fluid variables.  (The
scheme for indexing the variables is debatable, but the choice made
here is convenient for the present purposes.)
\begin{equation}
  \rho=\rho_0(r,\zeta,\tau)+\epsilon^2\rho_2(r,\zeta,\tau)+O(\epsilon^4),
\end{equation}
\begin{equation}
  p=\epsilon^2\left[p_0(r,\zeta,\tau)+\epsilon^2p_2(r,\zeta,\tau)+O(\epsilon^4)\right],
\end{equation}
\begin{equation}
  \Phi=\Phi_0(r)+\epsilon^2\Phi_2(r)\half\zeta^2+O(\epsilon^4),
\end{equation}
\begin{equation}
  v_r=\epsilon^2v_{r2}(r,\zeta,\tau)+\epsilon^4v_{r4}(r,\zeta,\tau)+O(\epsilon^6),
\end{equation}
\begin{equation}
  v_\phi=r\Omega_0(r)+\epsilon^2v_{\phi2}(r,\zeta,\tau)+\epsilon^4v_{\phi4}(r,\zeta,\tau)+O(\epsilon^6),
\end{equation}
\begin{equation}
  v_z=\epsilon\left[\epsilon^2v_{z2}(r,\zeta,\tau)+\epsilon^4v_{z4}(r,\zeta,\tau)+O(\epsilon^6)\right],
\end{equation}
\begin{equation}
  \psi=\epsilon\left[\psi_0(r,\tau)+\epsilon^2\psi_2(r,\zeta,\tau)+O(\epsilon^4)\right],
\end{equation}
\begin{equation}
  B_r=\epsilon\left[\epsilon B_{r1}(r,\zeta,\tau)+\epsilon^3B_{r3}(r,\zeta,\tau)+O(\epsilon^5)\right],
\end{equation}
\begin{equation}
  B_\phi=\epsilon\left[\epsilon B_{\phi1}(r,\zeta,\tau)+\epsilon^3B_{\phi3}(r,\zeta,\tau)+O(\epsilon^5)\right],
\end{equation}
\begin{equation}
  B_z=\epsilon\left[B_{z0}(r,\tau)+\epsilon^2B_{z2}(r,\zeta,\tau)+O(\epsilon^4)\right],
\end{equation}
\begin{equation}
  \nu=\epsilon^2\left[\nu_0(r,\zeta,\tau)+O(\epsilon^2)\right],
\end{equation}
\begin{equation}
  \eta=\epsilon^2\left[\eta_0(r,\zeta,\tau)+O(\epsilon^2)\right].
\end{equation}
We make the following observations.  The overall scaling of the
density is irrelevant provided that self-gravity is neglected;
however, the relative scaling of the pressure and density is
significant and implies that the sound speed is $O(\epsilon)$ (in
comparison to the orbital velocity).  The expansion of $\Phi$ is the
Taylor expansion of a smooth, symmetric external potential about the
midplane $z=0$; in the case of a point-mass potential
$\Phi=-GM(r^2+z^2)^{-1/2}$, we have $\Phi_0=-GM/r$ and
$\Phi_2=GM/r^3$.  The scaling of $\bB$ is such that, as mentioned
above, the magnetic field is dominated by the vertical component,
which has a magnetic pressure comparable to the gas pressure.  The
expansion of $\psi$ is of the form required to produce this.  The
scalings of $\nu$ and $\eta$ correspond to having (formally)
$\alpha=O(1)$.

Substituting these expansions into the basic equations and comparing
the coefficients of powers of $\epsilon$, we obtain a succession of
equations.  The radial component of the equation of motion at leading
order yields
\begin{equation}
  -\rho_0r\Omega_0^2=-\rho_0\p_r\Phi_0,
\end{equation}
which implies $r\Omega_0^2=\p_r\Phi_0$, i.e.\ that $\Omega_0(r)$ is
the angular velocity of a circular particle orbit of radius $r$ in the
midplane.  In the case of a point-mass potential, we obtain the
Keplerian value $\Omega_0=(GM/r^3)^{1/2}$.

The vertical component of the equation of motion at leading order yields
\begin{equation}
  0=-\rho_0\Phi_2\zeta-\p_\zeta p_0,
\end{equation}
which is the usual condition of hydrostatic equilibrium in which the
vertical pressure gradient balances the vertical gravitational force.
Under the present scalings, the Lorentz force does not affect this
balance at leading order.  The solution of this equation depends on
additional assumptions regarding the thermal physics of the disc.  In
the simplest situation of a vertically isothermal disc in which
$p_0=c_{\rms0}^2\rho_0$, where the isothermal sound speed
$c_{\rms0}(r,\tau)$ does not depend on $\zeta$, we obtain the familiar
solution
\begin{equation}
  \rho_0=\f{\Sigma_0}{(2\pi)^{1/2}H_0}\exp\left(-\f{\zeta^2}{2H_0^2}\right),
  	\label{density}
\end{equation}
where $H_0(r,\tau)=c_{\rms0}/\Phi_2^{1/2}$ is the isothermal
scaleheight and $\Sigma_0(r,\tau)$ is the surface density.

Next, the horizontal components of the equation of motion and of the
induction equation yield the four equations
\begin{eqnarray}
  \lefteqn{-2\rho_0\Omega_0v_{\phi2}=-\rho_0\p_r\Phi_2\half\zeta^2-\p_r\left(p_0+\f{B_{z0}^2}{2\mu_0}\right)}&\nonumber\\
  &&+\f{B_{z0}}{\mu_0}\p_\zeta B_{r1}+\p_\zeta(\rho_0\nu_0\p_\zeta v_{r2}),  \label{motion_r}
\end{eqnarray}
\begin{eqnarray}
  \lefteqn{\rho_0v_{r2}\f{1}{r}\p_r(r^2\Omega_0)=\f{B_{z0}}{\mu_0}\p_\zeta B_{\phi1}+\f{1}{r^2}\p_r(\rho_0\nu_0r^3\p_r\Omega_0)}&\nonumber\\
  &&+\p_\zeta(\rho_0\nu_0\p_\zeta v_{\phi2}),  \label{motion_phi}
\end{eqnarray}
\begin{equation}
  0=B_{z0}\p_\zeta v_{r2}+\p_\zeta[\eta_0(\p_\zeta B_{r1}-\p_rB_{z0})],
	\label{induction_br}
\end{equation}
\begin{equation}
  0=B_{r1}r\p_r\Omega_0+B_{z0}\p_\zeta v_{\phi2}+\p_\zeta(\eta_0\p_\zeta B_{\phi1}).
  		\label{induction_bphi}
\end{equation}
These can be regarded as four \emph{linear} equations for the unknowns
$v_{r2}$, $v_{\phi2}$, $B_{r1}$ and $B_{\phi1}$.  They are linear
because of the assumption of small deviations from orbital motion and
a vertical magnetic field, implicit in the asymptotic expansion.
Although these four quantities depend on $r$, $\zeta$ and $\tau$, only
derivatives with respect to $\zeta$ appear in these equations and they
can therefore be regarded an eighth-order system of ordinary
differential equations (ODEs) in $\zeta$ at each $r$ and $\tau$.  This
situation arises because the vertical dependence is assumed to be more
rapid than the radial dependence in a thin disc, and the
time-dependence is assumed to be slow.  The equations are
inhomogeneous and are driven by various source terms: the vertical
dependence of the radial gravitational force, the radial gradient of
total pressure, the azimuthal viscous force resulting from the orbital
motion, etc.

The expected symmetry of the solutions is that $v_{r2}$ and $v_{\phi2}$ are even functions of $\zeta$ while $B_{r1}$ and $B_{\phi1}$ are odd.  The behaviour of the solutions at large $|\zeta|$ can be illustrated by
considering the case of a vertically isothermal disc in which $\nu_0$
and $\eta_0$ are bounded as $|\zeta|\to\infty$ (e.g.\ if they are independent of $\zeta$).  In this case the solution is
of the form
\begin{equation}
  v_{r2}\sim C_1+E_1(\zeta), \label{ur_infinity}
\end{equation}
\begin{equation}
  v_{\phi2}\sim C_2\half\zeta^2\pm C_3\zeta+C_4+E_2(\zeta), \label{uphi_infinity}
\end{equation}
\begin{equation}
  B_{r1}\sim C_5\zeta\pm C_6+E_3(\zeta), \label{Br_infinity}
\end{equation}
\begin{equation}
  B_{\phi1}\sim \pm C_7 + E_4(\zeta) \label{Bphi_infinity}
\end{equation}
as $\zeta\to\pm\infty$, where the $C_i$ are constants and the $E_i$
are (like the density and pressure) exponentially small; here the
parametric dependence of the solution on $r$ and $\tau$ has been
suppressed.  In order to satisfy the governing equations, the values of the constants are
constrained by
\begin{equation}
  C_2B_{z0}=-C_5r\p_r\Omega_0,
\end{equation}
\begin{equation}
  C_3B_{z0}=-C_6r\p_r\Omega_0,
\end{equation}
\begin{equation}
  C_5=\p_rB_{z0};
\end{equation}
while $C_6$ and $C_7$ have the nature of input parameters, $C_1$ and $C_4$ have
the nature of output parameters.  The $C_5$ term is required in order
to produce a force-free field in the low-density exterior at large
$|\zeta|$.  (In fact it is a current-free or potential field, because an
axisymmetric poloidal force-free field must be current-free.) The
$C_6$ term represents, in some sense, the inclination of the poloidal
field at the upper surface of the disc, and we call
$C_6=B_{r\rms}$ accordingly.  More precisely, it is the radial
component of the magnetic field obtained by extrapolating the
parabolic form of the field lines in the force-free region above the
disc down to the midplane $\zeta=0$. The $C_7$ term represents the azimuthal component of the magnetic field at the upper surface and we call it $B_{\phi\rms}$. If we are considering discs without outflows (because of the condition $|B_{r\rms}|\ll|B_z|$) then we should set $B_{\phi\rms}=0$ in order that the magnetic stress $B_\phi B_z/\mu_0$ vanish at large $|\zeta|$. However, we retain this term as a convenient way to model the effect of angular momentum removal by an outflow without the complications of solving for the acceleration of the flow.  Angular momentum is removed by specifying a negative value for $B_{\phi\rms}B_z$.

In terms of the flux function $\psi$, we also have the relations
\begin{equation}
  B_{z0}=\f{1}{r}\p_r\psi_0,
\end{equation}
\begin{equation}
  B_{r1}=-\f{1}{r}\p_\zeta\psi_2.
\end{equation}
Neither $\psi_0$ nor $B_{z0}$ depends on $\zeta$, so $B_{z0}$ can be
regarded as an input parameter in the above system of ODEs.

Furthermore, the integrated form of the poloidal part of the induction
equation (equation~\ref{dpsidt}) yields
\begin{equation}
  \p_\tau\psi_0+v_{r2}\p_r\psi_0=\eta_0\left[r\p_r\left(\f{1}{r}\p_r\psi_0\right)+\p_\zeta^2\psi_2\right].
\label{psidot}
\end{equation}
Note that $-(1/r)\p_\zeta$ of equation~(\ref{psidot}) produces
equation~(\ref{induction_br}).  Once we have solved the system of
ODEs, equation~(\ref{psidot}) can be evaluated at any value of $\zeta$
(or indeed averaged in any convenient way with respect to $\zeta$) to
discover the value of $\p_\tau\psi_0$.  This gives the rate of
poloidal magnetic flux transport at the given values of $r$ and
$\tau$, which we can associate with an effective transport velocity
$v_\psi$ through $\p_\tau\psi_0+v_\psi\p_r\psi_0=0$, which implies
$\p_\tau B_{z0}+(1/r)\p_r(rv_\psi B_{z0})=0$.  In fact, if we evaluate
equation~(\ref{psidot}) in the limit $\zeta\to\infty$, we find
$\p_\tau\psi_0=-C_1rB_{z0}$ and therefore $v_\psi=C_1$.  Since the
electric current vanishes in this limit, the radial velocity
corresponds to the speed at which the magnetic field is transported.

In addition, the equation of mass conservation yields
\begin{equation}
  \p_\tau\rho_0+\f{1}{r}\p_r(r\rho_0v_{r2})+\p_\zeta(\rho_0v_{z2})=0.
\label{drhodt}
\end{equation}
We define the surface density at leading order,
\begin{equation}
  \Sigma_0(r,\tau)=\int_{-\infty}^\infty\rho_0(r,\zeta,\tau)\,\rmd\zeta.
\end{equation}
Then (assuming no vertical mass loss from the disc)
\begin{equation}
  \p_\tau\Sigma_0+\f{1}{r}\p_r\left(r\int_{-\infty}^\infty\rho_0v_{r2}\,\rmd\zeta\right)=0.
\end{equation}
Let
\begin{equation}
  m_0(r,\tau)=\int_0^r\Sigma_0(r',\tau)\,2\pi r'\,\rmd r'
\end{equation}
be the mass contained within radius $r$ at time $\tau$; then $\p_\tau
m_0+v_m\p_rm_0=0$, where
\begin{equation}
  v_m=\f{1}{\Sigma_0}\int_{-\infty}^\infty\rho_0v_{r2}\,\rmd\zeta
\end{equation}
is the effective transport velocity for mass, which can be determined
from the solution of the system of ODEs.

In summary, then, the input parameters at given values of $r$ and
$\tau$ are $\Omega_0$, $\Phi_2$, $c_{\rms0}$, $\Sigma_0$, $\nu_0$,
$\eta_0$, $B_{z0}$, $B_{r\rms}$, $B_{\phi\rms}$, $\p_r\Omega_0$, $\p_r^2\Omega_0$,
$\p_rc_{\rms0}$, $\p_r\Sigma_0$, $\p_r\nu_0$, $\p_r\eta_0$ and $\p_rB_{z0}$.  The output parameters of greatest interest are $v_m$
and $v_\psi$.  The problem can be simplified, and made dimensionless,
by assuming a Keplerian disc ($\Omega_0\propto r^{-3/2}$ and
$\Phi_2=\Omega_0^2$) and alpha prescriptions of the form $\nu_0=\alpha
c_{\rms0}^2/\Omega_0$, $\eta_0=\nu_0/\Pm$ with $\alpha$ and $\Pm$
being constants.

In this analysis we have simplified the problem by assuming that the
magnetic field is dominated by its vertical component within the disc.
This field needs to be matched to a force-free (in fact, current-free)
field outside the disc, which involves the solution of a global
problem \citep[e.g.][]{ogilvie97}.  The nature of this problem is that
the poloidal flux distribution on the disc determines the inclination
of the poloidal field at each radial location, and in general this
angle is not small.  An alternative ordering scheme is possible, in
which all three components of the magnetic field are $O(\epsilon)$.
In such a scheme, the inclination of the poloidal field could be large
and magneto-centrifugal jet launching would be possible
\citep{ogilvie01}.  The pressure of the horizontal magnetic field
would also contribute to the vertical force balance and the problem
would become more nonlinear.  (In fact, it is easy to add this effect
by hand to the present system of equations, but we will not do so
here.)  However, the transport of magnetic flux by diffusion would be
formally faster by a factor $O(\epsilon^{-1})$ than that by advection,
if $\Pm=O(1)$; this inequality is merely an expression of the result
of \citet{lubow94a}.  Therefore no ordering scheme is fully
satisfactory.  We adopt the present scheme because of the transparency
of its derivation, the linearity of the resulting equations, and
because it avoids the complications of jet launching, while missing
only the effect of the magnetic compression of the disc.

\section{A quasi-local problem}
\label{sec:local}

Equations~(\ref{motion_r})--(\ref{induction_bphi}) define a
quasi-local problem, as only vertical derivatives of the unknowns
appear, while radial derivatives appear only as source terms. The
solution of this local problem provides the radial transport
velocities of mass and poloidal magnetic flux (at a given time and
radius), which are necessary to determine the global evolution of the
disc on a viscous/resistive timescale. The local problem on the other
hand is effectively stationary (no time derivatives appear), because
stationarity is assumed on the much shorter dynamical timescale. In
this article and paper II, we solve this local problem and leave for future
work the study of the global evolution of a disc. In this section we
rewrite in a dimensionless form the equations governing the local
problem. We discuss the dependence on the different parameters, and
describe the numerical method used to solve the equations.

\subsection{Differential system}
\label{sec:system}
We choose as independent variables $\rho_0$, $v_{r2}$, $v_{\phi1}$,
$B_{r1}$ and $B_{\phi1}$. These variables are non-dimensionalised by
defining
\begin{eqnarray}
\tilde{\rho} &\equiv & \frac{\rho_0 H}{\Sigma}, \\
u_r &\equiv & \frac{r}{H}\frac{v_{r2}}{c_\rms}, \\
u_\phi &\equiv &  \frac{r}{H}\frac{v_{\phi2}}{c_\rms}, \\
b_r &\equiv &  \frac{r}{H}\frac{B_{r1}}{B_z}, \\
b_\phi &\equiv &  \frac{r}{H}\frac{B_{\phi1}}{B_z}, 
\end{eqnarray}
where we have dropped the subscript $0$ on quantities such as $H$,
$c_\rms$, $B_z$ and $\Sigma$. A dimensionless vertical spatial coordinate is also defined as: 
\begin{equation}
\zeta \equiv z/H.
\end{equation} 
Note that this variable is different from $\zeta$ used in Section~\ref{sec:asymptotic}, which was a rescaled but dimensional variable. In the rest of the paper when we use the symbol $\zeta$, we refer to the dimensionless variable. 

We assume a point-mass potential and therefore circular Keplerian
orbital motion at leading order.  We use the standard alpha
prescription for the viscosity, $\nu = \alpha c_\rms^2/\Omega = \alpha
c_\rms H$, allowing for a vertical (but not radial) dependence of the
$\alpha$ parameter through $\alpha = \alpha_0g(\zeta)$. The
resistivity $\eta$ is then related to the viscosity through the
magnetic Prandtl number $\Pm \equiv \nu/\eta $, which can also have a vertical dependence. An
isothermal equation of state is assumed for simplicity.
Equation~(\ref{density}) gives the corresponding density profile,
which can be written in dimensionless form as
\begin{equation}
\tilde{\rho} = \frac{1}{\sqrt{2\pi}}\,\mathrm{e}^{-\zeta^2/2}.
\end{equation}

The differential system given by equations~(\ref{motion_r})--(\ref{induction_bphi}) is rewritten as
\begin{eqnarray}
-\frac{1}{\tilde\rho}\lefteqn{\p_\zeta\left(\tilde{\rho}\alpha\p_\zeta u_r \right) - 2u_\phi - \frac{1}{\beta_0\tilde\rho} \p_\zeta b_r = \frac{3}{2} + D_H - D_{\nu\Sigma}}&\nonumber\\
&& + \left(\frac{3}{2}-D_H\right)\zeta^2 - \frac{D_B}{\beta_0\tilde\rho}, \label{eq:motion_r}
\end{eqnarray}
\begin{eqnarray}
-\frac{1}{\tilde\rho}\lefteqn{\p_\zeta\left(\tilde{\rho}\alpha\p_\zeta u_\phi \right) + \frac{1}{2} u_r - \frac{1}{\beta_0\tilde\rho}\p_\zeta b_\phi = \frac{3}{2}\alpha \bigg\lbrack - \frac{1}{2} }&\nonumber\\
&& + D_H\left(1 + \frac{\dd\ln\alpha}{\dd\ln\zeta} \right)- D_{\nu\Sigma} - D_H\zeta^2 \bigg\rbrack,  \label{eq:motion_phi}
\end{eqnarray}
\begin{equation}
-\p_\zeta\left( \frac{\alpha}{\Pm}\p_\zeta b_r \right) - \p_\zeta
u_r = -D_B \p_\zeta\left(\frac{\alpha}{\Pm}\right), \label{eq:induction_r}
\end{equation}
\begin{equation}
 -\p_\zeta\left(\frac{\alpha}{\Pm}\p_\zeta b_\phi\right) - \p_\zeta u_\phi + \frac{3}{2}b_r = 0,\label{eq:induction_phi}
\end{equation}
where we have defined the following dimensionless parameters:
\begin{eqnarray}
\beta_0 &\equiv& \frac{\mu_0}{B_z^2}\frac{\Sigma c_s^2}{H}, \\
D_H &\equiv&\frac{\p\ln H}{\p\ln r}, \\
D_{\nu\Sigma} &\equiv& 2D_H - \frac{3}{2} + \frac{\p\ln\Sigma}{\p\ln r}, \\
D_B &\equiv & \frac{\p\ln B_z}{\p\ln r}.
\end{eqnarray}
Here $\beta_0$ corresponds roughly to the midplane value of the plasma $\beta$ parameter (the ratio of the thermal pressure to the magnetic pressure); more precisely, the two are related by $\beta(\zeta=0) = \sqrt{2/\pi}\,\beta_0 $. The parameter $D_{\nu\Sigma}$ equals $\p\ln(\nu\Sigma)/\p\ln r$, given that $\nu=\alpha H^2\Omega$ and $\alpha$ does not depend on $r$.

\subsection{Boundary conditions}
From equations~(\ref{ur_infinity})--(\ref{Bphi_infinity}), one can
deduce that the following quantities vanish exponentially fast as
$\zeta \rightarrow \pm \infty$:
\begin{equation}
\rho u_r \to0, \label{eq:boundary_ur} \\
\end{equation}
\begin{equation}
\rho u_\phi \to0, \label{eq:boundary_uphi}
\end{equation}
\begin{equation}
b_r-(D_B\zeta\pm b_{r\rms})\to0, \label{eq:boundary_br}
\end{equation}
\begin{equation}
b_\phi - (\pm b_{\phi\rms}) \to 0. \label{eq:boundary_bphi} \\
\end{equation}
Let us recall that these conditions are motivated by the absence of
outflow, owing to the low inclination of the field lines. As a
consequence the magnetic field at infinity is force-free. The inclusion of a non-vanishing azimuthal component of the magnetic field is not self-consistent but allows us, if we wish, to mimic the effect of angular momentum removal by an outflow. These boundary conditions are homogeneous,
except for the linear source terms proportional to $D_B$, $b_{r\rms}$, and $b_{\phi\rms}$ in equations~(\ref{eq:boundary_br}) and (\ref{eq:boundary_bphi}).

As the differential equations, the source terms and boundary
conditions have reflectional symmetry about the midplane, the
stationary profile we seek shares the same symmetry (implying a
symmetric horizontal velocity and an antisymmetric horizontal magnetic
field). As a consequence, the following conditions apply at the
midplane $\zeta=0$:
\begin{equation}
\p_\zeta u_r=0, \label{eq:midplane_ur}
\end{equation}
\begin{equation}
\p_\zeta u_\phi=0,  \label{eq:midplane_uphi}
\end{equation}
\begin{equation}
b_r=0,  \label{eq:midplane_br}
\end{equation}
\begin{equation}
b_\phi=0.  \label{eq:midplane_bphi}
\end{equation}
Reflectional symmetry also implies that the solution need be computed
only in the half-space $\zeta>0$.

\subsection{Transport velocities of mass and magnetic flux}
The transport rates of mass and magnetic flux can be obtained from the solution of the above equations and expressed in terms of a transport velocity in the following way. Multiplying the azimuthal component~(\ref{eq:motion_phi}) of the equation of motion by $2\tilde\rho$, integrating over all $\zeta$ and
applying the boundary conditions, we obtain
\begin{equation}
  \int_{-\infty}^\infty\tilde\rho u_r\,\rmd\zeta=-\f{3}{2}(1+2D_{\nu\Sigma})\int_{-\infty}^\infty\tilde\rho\alpha\,\rmd\zeta + \frac{4}{\beta_0}b_{\phi\rms}.
\end{equation}
In the case of uniform $\alpha$ this gives the transport velocity
\begin{equation}
  u_m=-\f{3}{2}\alpha(1+2D_{\nu\Sigma}) + \frac{4}{\beta_0}b_{\phi\rms},
\label{um}
\end{equation}
where
\begin{equation}
  u_m\equiv\f{r}{H}\f{v_m}{c_\rms}=\int_{-\infty}^\infty\tilde\rho u_r\,\rmd\zeta
\end{equation}
is the dimensionless mass transport velocity.  More generally,
$\alpha$ should be replaced by $\alpha_0$ in equation~(\ref{um}) and a
factor of
\begin{equation}
  \int_{-\infty}^\infty\rme^{-\zeta^2/2}g(\zeta)\,\rmd\zeta\Bigg/\int_{-\infty}^\infty\rme^{-\zeta^2/2}\,\rmd\zeta
\end{equation}
should be included. For a vanishing $b_{\phi\rms}$ these results are consistent with the familiar
expression from the standard theory of a Keplerian accretion disc,
\begin{equation}
  \bar v_r=-\f{3}{r^{1/2}\Sigma}\p_r(r^{1/2}\bar\nu\Sigma),
\end{equation}
where the overbar refers to a density-weighted vertical average.

The integrated version of the radial component~(\ref{eq:induction_r})
of the induction equation is
\begin{equation}
u_\psi = u_r + \frac{\alpha}{\Pm}(\p_\zeta b_r-D_B) = \mathrm{const},
\label{upsi}
\end{equation}
where
\begin{equation}
  u_\psi\equiv\f{r}{H}\f{v_\psi}{c_\rms}
\end{equation}
is the dimensionless magnetic flux transport velocity. It can be evaluated using equation~(\ref{upsi}) at any convenient hight.

\subsection{Relation to previous works}

\citet{ogilvie01} solved for the vertical structure of a magnetised
accretion disc, allowing for the possibility of magneto-centrifugal jet
launching \citep[see also][]{ogilvie97,ogilvie98a}.  The equations they
solved are similar to ours, but with the following differences.  The
thermal structure of the disc was determined by a balance between
viscous and resistive heating and radiative cooling, rather than being
assumed to be isothermal.  This solution was matched to an isothermal
atmosphere with a force-free magnetic field.  The radial gradients of
thermal and magnetic pressure were neglected, as was the vertical
transport of momentum by viscosity.  The pressure of the horizontal
magnetic field was taken into account in the vertical force balance.

\citet{lovelace09} and \citet{bisnovatyi-kogan12} also performed a similar calculation of the vertical structure of a magnetised accretion disc. The differences between their approach and ours are the following. The equations in the limit of a thin disc were derived more systematically in this paper, and describe more precisely the vertical structure: contrary to \citet{lovelace09} we take into account the vertical dependence of the pressure gradient and viscous stress. We also did not discard the radial gradient of vertical magnetic field, whose contribution to the magnetic flux transport appears at the same order in the expansion as the advection by an accretion flow. Furthermore, we solve a more general problem, as we do not assume that the disc is stationary on a viscous/resistive timescale but only on a dynamical timescale, and we consider a much wider range of $\beta$. The vertical boundary conditions also differ substantially. We calculate the vertical profiles up to the region that is magnetically dominated, where one can impose the magnetic field to be that dictated by the exterior solution. On the other hand, \citet{lovelace09} and \citet{bisnovatyi-kogan12} do not solve the transition between the interior of the disc and the magnetically dominated region; instead they apply special boundary conditions at the surface of the disc. The validity of these boundary conditions may be questioned in light of the results of Section~\ref{sec:cstalpha}, since we find a significant jump in $b_r$ and $v_\phi$ at the transition to the force-free regime, while their boundary conditions preclude such jumps (note however that their diffusion coefficients vanish in the surface layer, which is not considered in this article but delayed to paper II). Another difference is the choice of $\alpha$:  considering the transition to the magnetically dominated regime forces us to determine $\alpha$ through the marginal stability hypothesis in order to avoid unphysical effects (see Section~\ref{sec:marginal}). Finally, they use a disc model where the density is independent of height inside the disc, while we use an isothermal model where the density is a Gaussian function of height. Allowing a variation of the density is essential in our result that the advection speed of mass and magnetic flux can differ dramatically (see Section~\ref{sec:cstalpha}).

\subsection{Relation to the shearing sheet}

Both our equations and those of \citet{ogilvie01} are related to the
local approximation (shearing sheet or shearing box), commonly used in
the study of accretion discs.  Starting with the equations of
MHD in this approximation and assuming a solution
that depends only on the vertical coordinate $z$, we would obtain
equations equivalent to ours except that the source terms would be
absent and the pressure of the horizontal magnetic field would be
taken into account in the vertical force balance.

\subsection{Relation to the magnetorotational instability}

The four equations~(\ref{eq:motion_r})--(\ref{eq:induction_phi}) governing
the horizontal components of the velocity and magnetic field are
closely related to those appearing in the analysis of the
magnetorotational instability \citep[MRI;][]{balbus98}.  Let us write
the equations in the symbolic form
\begin{equation}
  \mathbf{L}\bmath{X}=\bmath{F},
  \label{lx}
\end{equation}
where $\bmath{X}=[u_r\;u_\phi\;b_r\;b_\phi]^\mathrm{T}$ is a vector of
unknowns, $\mathbf{L}$ is the linear operator that generates the
left-hand sides of the equations, and $\bmath{F}$ represents the source
terms on the right-hand sides.  To analyse the MRI of an isothermal
disc with a vertical magnetic field, we would instead solve the
eigenvalue problem
\begin{equation}
  \mathbf{L}\bmath{X}=\lambda\bmath{X},
\label{lx_mri}
\end{equation}
with homogeneous boundary conditions, corresponding to solutions of
the linearized equations proportional to $\mathrm{e}^{\lambda t}$ and
with no horizontal spatial dependence.  The MRI would correspond to a
solution with $\mathrm{Re}(\lambda)>0$.  The close connection between
the MRI and the equilibrium of magnetised discs has been noted before
\citep{ogilvie98b,ogilvie01}.

\subsection{Marginal stability hypothesis}
	\label{sec:marginal}

The value of $\alpha$ (or $\alpha_0$ if a vertical dependence is
permitted) can be estimated by supposing that the MRI is made
marginally stable by the effect of the viscosity and resistivity
\citep{ogilvie01}. This approach has the desirable property of
ensuring that the obtained stationary solution is not unstable (at
least to modes with no horizontal dependence, which are invariably the
most dangerous).  \citet{ogilvie01} showed that this assumption avoids
unphysical effects like multiple bending of the field lines (see also Appendix~A).

\begin{figure*}
   \centering
   \includegraphics[width= 12cm]{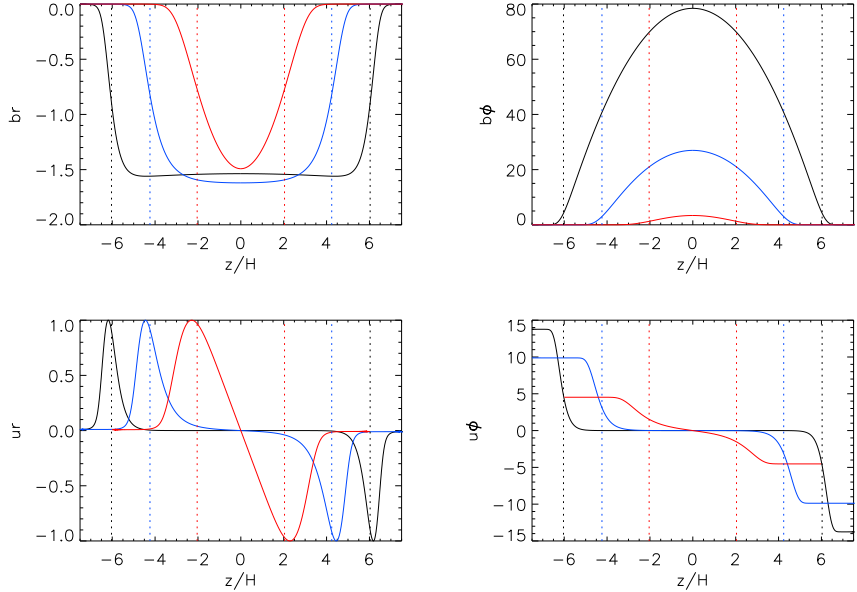}
   \caption{Marginal MRI channel modes for different strength of the vertical magnetic field : $\beta_0 =  10$ (red), $\beta_0 =10^4$ (blue), and $\beta_0 =10^8$ (black). The calculation is done with uniform diffusion coefficients and $\Pm = 1$. The modes are normalised such that their maximum radial velocity equals one. The vertical dotted lines illustrate the height $\zeta_B$ at which the magnetic and thermal pressures coincide, and where the marginal channel modes tend to localise.}
   \label{fig:channel}
\end{figure*}

Stability depends on the strength of the magnetic field (through
$\beta_0$) and on the viscosity and resistivity (through $\alpha$ and
$\Pm$).  As expected, the most difficult mode to stabilise is the
largest-scale channel mode, which is antisymmetric about the midplane
\citep{ogilvie01}, and this mode has $\lambda=0$ at marginal
stability.  \footnote{Although this means that the linear operator $\mathbf{L}$
is singular and possesses an antisymmetric null eigenfunction,
equation~(\ref{lx}) still has a unique solution that is symmetric about
the midplane, because the operator is invertible within the symmetric
subspace.}  For weak fields ($\beta_0 \ga 10^4$) the marginal mode is
localised around $\zeta =\pm\zeta_B$ (Figure~\ref{fig:channel}), where 
\begin{equation}
\zeta_B = \sqrt{\ln\left(\frac{2}{\pi}\beta_0^2\right)} 
	\label{eq:zetab}
\end{equation}
is the height at which
the magnetic pressure equals the thermal pressure.

\begin{figure}
   \centering
   \includegraphics[width=\columnwidth]{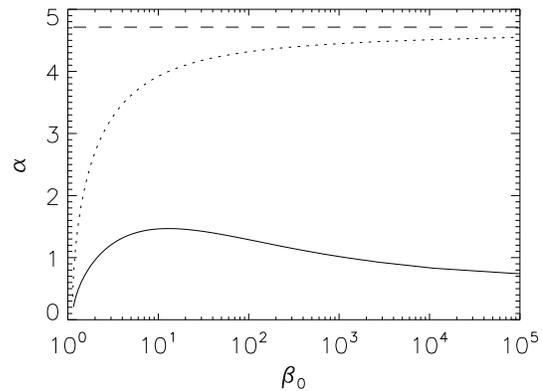}
   \caption{The value of the $\alpha$ parameter determined by the marginal stability hypothesis as a function of the magnetic field strength. The calculation is done with uniform diffusion coefficients. The full line illustrates the value of $\alpha$ obtained with the magnetic Prandtl number mostly used in this paper, $\Pm=1$. The dotted line shows the $\alpha$ values obtained without viscosity ($\Pm=0$), and the dashed line the corresponding analytical estimate for large $\beta_0$ of Appendix~A (equation~\ref{eq:alpha_marginal}). }
   \label{fig:alpha}
\end{figure}
 
While physically motivated, this way of determining the effective
turbulent diffusion coefficients should be taken with a grain of
salt. Particularly worrying is the tendency of this method to produce
large values of $\alpha$. Indeed, in the case of uniform diffusion
coefficients, $\alpha$ is of order unity over a large range of
magnetic field strength (Figure~\ref{fig:alpha}). Such a high value is
most probably not realistic for weak magnetic fields ($\beta_0 \gg
1$), although we note that direct numerical simulations of discs with
vertical gravity and a net vertical magnetic flux are computationally
difficult and few results are available, perhaps precisely because the
turbulence is very intense when $\beta_0$ is not very large. The
somewhat surprisingly slow decline of $\alpha$ at large $\beta_0$ is
due to the fact that its value is determined mostly by the region
where the channel mode is localised, which is rather strongly
magnetised ($\beta\sim1$) even when the midplane is very weakly
magnetised. Obviously, assuming the same value of $\alpha$ in these
regions with vastly different magnetisation is unrealistic. The use of
a vertical profile for $\alpha$, as is done in paper II, alleviates this issue by reducing the midplane value ($\alpha(\zeta=0)
\sim 0.1$ is more realistic although still probably too high for very
weak fields), while keeping similar values far from the
midplane.

Fortunately (given the uncertainty on the value of $\alpha$), the
balance between the advection and diffusion of the magnetic field is
only weakly dependent on $\alpha$, when the angular transport enabling advection is caused by turbulence. Indeed, both advection and diffusion processes are (at least approximately) proportional to $\alpha$, through respectively the viscosity
and the resistivity.

\subsection{Parameters and form of the solution}
\label{sec:parameters}
The free parameters of the dimensionless problem are the magnetic Prandtl number
$\Pm$ and the magnetisation parameter $\beta_0$, as well as the source
terms $D_H$, $D_{\nu\Sigma}$, $D_B$, $b_{r\rms}$ and  $b_{\phi\rms}$.

An important consequence of the linearity of the equations is that the solution depends linearly on the source terms, appearing either on the right-hand side of the system of equations or as a non-vanishing boundary condition at infinity ($b_{r\rms}$ and  $b_{\phi\rms}$). Thus, the general solution is a linear combination of the solution vectors corresponding to each source term:
\begin{eqnarray}
\lefteqn{\bmath{X} = \bmath{X}_\mathrm{K} + \bmath{X}_{DH}D_H + \bmath{X}_{D\nu\Sigma}D_{\nu\Sigma} + \bmath{X}_{DB}D_B}&\nonumber\\
&& + \bmath{X}_{br\rms}b_{r\rms} + \bmath{X}_{b\phi\rms}b_{\phi\rms},
\end{eqnarray}
where $\bmath{X}_{DH}$ is the solution vector corresponding to the source term proportional to $D_H$ and so on. $\bmath{X}_\mathrm{K}$ is the solution vector corresponding to the source terms $\bmath{F}$ that are independent of
$D_H$, $D_{\nu\Sigma}$, $D_B$, $b_{r\rms}$, and $b_{\phi\rms}$; these terms contain
the radial derivative of the leading-order angular velocity which is
assumed to be Keplerian (hence the notation), as well as a term describing the vertical dependence of the gravitational potential and other geometrical terms coming from differentiating factors depending on $r$ in the viscous term and the radial pressure gradient. Furthermore, we define
\begin{equation}
\bmath{X}_\mathrm{hyd} = \bmath{X}_\mathrm{K} + \bmath{X}_{DH},
\end{equation}
being the solution corresponding to the hydrodynamic source terms with
the standard parameters $D_H=1$ (relevant to a disc with a constant
aspect ratio $H/r$) and $D_{\nu\Sigma}=0$ (relevant to a steady
accretion disc far from the inner boundary). For these parameters, we
thus have
\begin{equation}
\bmath{X} = \bmath{X}_\mathrm{hyd} +  \bmath{X}_{DB}D_B + \bmath{X}_{br\rms}b_{r\rms} +  \bmath{X}_{b\phi\rms}b_{\phi\rms}.
\end{equation}

The linearity of the problem makes the exploration of the parameter
space and the physical understanding of the solutions much easier.
Indeed, given the marginal stability hypothesis, the solution depends
in a nonlinear way only on the two parameters $\beta_0$ and $\Pm$. For
each pair of values of these two parameters, one needs to compute the
six solution vectors $\bmath{X}_\mathrm{K}$, $\bmath{X}_{DH}$, $
\bmath{X}_{D\nu\Sigma}$, $\bmath{X}_{DB}$,  $\bmath{X}_{br\rms}$ and $\bmath{X}_{b\phi\rms}$,
each of which can be represented by plotting the profiles of $u_r$,
$u_\phi$, $b_r$ and $b_\phi$.  The general solution is then just a
linear combination of these solution vectors with the appropriate
coefficients.

Note that another input to the model is the vertical profile of the
diffusion coefficients, the shape of which can be freely imposed
(although its normalisation is determined by the marginal stability
hypothesis).  In Section~\ref{sec:cstalpha}, we study in detail the
simplest case of a uniform resistivity and viscosity. The effect of the vertical
structure of the diffusion coefficients will be considered in paper II.

\subsection{Method of numerical solution}

\begin{figure*} 
   \centering
   \includegraphics[width= 12cm]{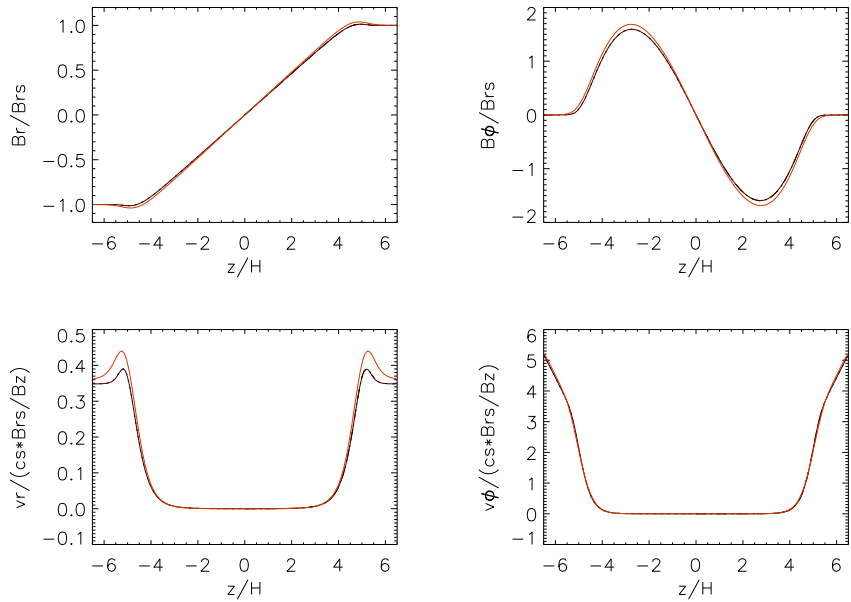}
   \caption{Comparison of 1D time-dependent direct numerical simulations of a stratified shearing box (full lines) and solutions by the spectral method (dashed line) and the shooting method (dotted line). The black full line is a simulation with $B_{r\rms}=10^{-3}B_z$ (it is almost indistinguishable from the dashed and dotted lines), while the orange line is a simulation with $B_{r\rms}=0.5B_z$. Other parameters are $\beta_0=10^4$, $\alpha=1.5$, $\Pm=1$, and all source terms except $b_{r\rms}$ are set to zero.}
   \label{fig:simu}
 \end{figure*}

We solve the problem described in the previous subsections with the
use of two different methods: the shooting method, and a (more successful) spectral
method using a decomposition on a basis of Whittaker cardinal
functions, which are well suited to problems on an infinite interval
\citep[e.g.][]{boyd01,latter10}. In the shooting method, we shoot from
the midplane by guessing the values of $u_r$, $u_\phi$, $\p_\zeta
b_r$, and $\p_\zeta b_\phi$ there (equations~\ref{eq:midplane_ur}--\ref{eq:midplane_bphi}). The differential system given by equations~(\ref{eq:motion_r})--(\ref{eq:induction_phi}) is then
used to evolve this solution up to a height $\zeta_{\rm num}$, where
the boundary conditions at infinity are applied. Newton iteration is
used to converge to the desired solution. $\zeta_{\rm num}$ should be
chosen large enough to lie in the force-free regime, in which case the solution is independent of $\zeta_{\rm num}$. We found that choosing
$\zeta_{\rm num}$ typically one or two scale-heights above $\zeta_B$
was sufficient for this purpose.  Similarly, the `grid' of the
spectral method should extend up to several scale-heights above
$\zeta_B$ to obtain converged results. The decomposition on Whittaker
cardinal functions implicitly imposes the condition that the variables
tend to zero exponentially fast at infinity. For this reason, in the
numerical calculation using the spectral method we replace $b_r$ and $b_\phi$ by
the following variables:
\begin{equation}
\tilde{b}_r \equiv b_r - D_B\zeta - b_{r\rms}\tanh(\zeta^3),
\end{equation} 
\begin{equation}
\tilde{b}_\phi \equiv b_\phi - b_{\phi\rms}\tanh(\zeta^3),
\end{equation} 
which should vanish exponentially fast at infinity according to the
boundary conditions stated in equations~(\ref{eq:boundary_br})--(\ref{eq:boundary_bphi}). The
differential system had to be changed accordingly. We obtained very
good agreement between the two methods (Figure~\ref{fig:simu}).

We also compared the results (with no source terms except $b_{r\rms}$) with one-dimensional time-dependent direct numerical simulations of a stratified shearing box in which the value of $B_r$ is imposed at the vertical boundaries. The simulations were performed with the code RAMSES \citep{teyssier02,fromang06}. The comparison allows us to evaluate the limitation due to our assumption that $|B_r| \ll |B_z|$ (Figure~\ref{fig:simu}). Fortunately even when the field is significantly inclined ($B_r =0.5 B_z$, just below the inclination threshold above which a magnetocentrifugal jet is launched), the velocity and magnetic field profiles remain very close to those obtained under the assumption of a small inclination. This is observed to be true as long as $\beta_0$ is rather large. For $\beta_0$ values of order unity magnetic compression (neglected here) would be expected to play a significant role if the inclination is large.

The results described in the remainder of this paper were obtained with the spectral method, which gave a good convergence on a broader parameter space than the shooting method.

\section{The case of uniform diffusion coefficients}
	\label{sec:cstalpha}

The formalism developed in Sections~\ref{sec:asymptotic} and \ref{sec:local} allows in principle for a vertical dependence of the diffusion coefficients. However, in the remainder of this paper we focus on the case of uniform diffusion coefficients. This simple, but probably unrealistic, model is a first step and has the advantage of allowing for some analytical treatment. More general models involving non-uniform diffusion coefficients are studied in paper II. In the remainder of this section, unless otherwise noted, the numerical calculations use our fiducial parameters $\Pm=1$ and $\beta_0=10^4$.

\subsection{Approximate analytical model}
      \label{sec:anal}

Before discussing the numerical solutions of our problem, we develop an approximate analytical model that is helpful in their interpretation. The magnetic field and velocity profiles can be easily understood in the two limiting cases of
weak (passive) magnetic field ($\beta \gg 1$), and very strong (force-free) magnetic field ($\beta \ll 1$). In the case where the magnetic
pressure is small compared to the midplane pressure ($\beta_0 \gg 1$),
both regimes appear: close to the midplane the field is passive, while
at $\zeta \rightarrow \infty$ the field is force-free. The transition
between the two regimes takes place around $\zeta_B$ where the
magnetic pressure is comparable to the thermal pressure ($\zeta_B$ is given by equation~(\ref{eq:zetab})). In the
following two subsections we describe the two different regimes. In a
third subsection we construct a simple model that connects the
two regions to obtain an analytical estimate of the profiles, and in
particular of the transport velocity of the magnetic flux.

\subsubsection{Passive magnetic field}
     \label{sec:passive_field}
In the limit of very large $\beta$, the Lorentz force is negligible
and the velocity profile is unaffected by the magnetic field.
If the kinematic viscosity is uniform, the purely hydrodynamic velocity profiles
can be computed analytically and are parabolic:
\begin{eqnarray}
u_r   &=& u_{r0} + u_{r2}\zeta^2, \label{eq:ur_hydro} \\
u_\phi &=& u_{\phi0} + u_{\phi2}\zeta^2, \label{eq:uphi_hydro} 
\end{eqnarray}
with the following coefficients:
\begin{eqnarray}
u_{r0} &=& \alpha\left(-\frac{9}{2} + 5D_H - 3D_{\nu\Sigma}\right) + \frac{4\alpha^3}{1+4\alpha^2}\left(3-5D_H\right), \nonumber\\ \\
u_{r2} &=& \frac{\alpha}{1+4\alpha^2}\left(3-5D_H\right),  \\
u_{\phi0} &=& -\f{1}{2}\left(D_H - D_{\nu\Sigma} + \frac{3}{2}\right) -  \frac{\alpha^2}{1+4\alpha^2}\left(3-5D_H\right), \nonumber\\  \\
u_{\phi2} &=& \f{1}{2}\left(D_H - \frac{3}{2}\right) +  \frac{\alpha^2}{1+4\alpha^2}\left(3-5D_H\right). \label{eq:uphi2}
\end{eqnarray}

The magnetic field profiles correspond then to a situation where the
stretching of the magnetic field lines by the (imposed) velocity field is compensated by the
diffusion following equations~(\ref{eq:induction_r})--(\ref{eq:induction_phi}). These can
be solved if some boundary conditions can be applied, which is not
obvious as the region where the magnetic field is passive does not
extend to infinity (see Section~\ref{sec:anal_model} for a simple
model). After two successive integrations, the radial induction equation gives
\begin{equation}
b_r(\zeta) = b_{r1}\zeta -\frac{\Pm}{\alpha}\frac{u_{r2}}{3}\zeta^3 , \label{eq:br_passive}
\end{equation}
where $b_{r1}$ is some unknown constant to be determined by the
boundary conditions, and we used $b_r(\zeta=0) = 0$ to remove one
integration constant. Similarly, the azimuthal induction equation gives
\begin{equation}
b_\phi(\zeta) = b_{\phi1}\zeta + \frac{\Pm}{\alpha}\left\lbrack \left(\frac{b_{r1}}{4}- \frac{u_{\phi2}}{3}\right)\zeta^3 -
\frac{\Pm}{\alpha}\frac{u_{r2}}{40}\zeta^5  \right\rbrack, \label{eq:bphi_passive}
\end{equation}
where $b_{\phi1}$ is to be determined by boundary conditions.

\subsubsection{Force-free magnetic field}
     \label{sec:force_free}
When $\beta \ll 1$, nothing can compensate the Lorentz force, so the
magnetic field has to be force-free: the current is parallel to the
magnetic field lines. With our assumption that the field is almost
vertical, this means that the radial and azimuthal currents vanish:
\begin{eqnarray}
 \p_\zeta b_\phi &=& 0, \\
\p_\zeta b_r - D_B &=& 0.
\end{eqnarray}
Applying the boundary conditions at infinity, we find that they are satisfied anywhere in the force-free
region:
\begin{eqnarray}
b_\phi &=& \pm b_{\phi\rms}, \\
b_r &=& \pm b_{r\rms} + D_B\zeta,
\end{eqnarray}
where $\pm$ stands for $\mathrm{sgn}(\zeta)$.

Because the current vanishes, the magnetic field cannot
diffuse: the velocity is dictated by the fact that the fluid is frozen
into the magnetic field lines. In particular, isorotation is enforced
along the magnetic field lines, determining the azimuthal velocity to be
\begin{equation}
u_\phi = \frac{3}{2}\left(b_{r\rms}|\zeta| + \frac{D_B}{2}\zeta^2 \right) + u_{\phi0}^{\prime},
\end{equation}
where $u_{\phi0}^{\prime}$ is again a constant to be determined by boundary conditions. The radial velocity is constant and equals the
speed $u_\psi$ at which the magnetic field is being advected or diffused.

\subsubsection{Two-zone model}
     \label{sec:anal_model}
In this subsection, we build an approximate model of the vertical profiles of velocity and magnetic field
by assuming that for $\zeta < \zeta_B$ they behave as described in
Section~\ref{sec:passive_field} (passive field) and for $\zeta >
\zeta_B$ they behave as in Section~\ref{sec:force_free} (force-free
field). By doing so we neglect the thickness of the transition where
$\beta$ is of order unity. To build the model, we need to connect the two regions of passive field
($\zeta < \zeta_B$) and force-free field ($\zeta > \zeta_B $), thus determining the proper boundary
conditions at $\zeta = \zeta_B$ (the conditions at $\zeta=-\zeta_B$ then being given simply by reflectional symmetry). Four conditions are needed to constrain the four unknowns $b_{r1}$, $b_{\phi1}$, $u_{\phi0}^\prime$ and $u_\psi$.

Two boundary conditions can be obtained from the analysis
of the induction equation. The radial component states that the
azimuthal electric field is independent of the height $\zeta$ (equation~\ref{upsi}). Using
equations~(\ref{eq:ur_hydro}) and~(\ref{eq:br_passive}) at $\zeta = 0$, we obtain a first condition:
\begin{equation}
 u_{\psi} = \frac{\alpha}{\Pm}\left( b_{r1} - D_B\right) + u_{r0}.  \label{eq:const_ur}
\end{equation}

The azimuthal component of the induction equation is more complicated because
$b_r$ acts as a source term in this equation. By integrating between
two heights $\zeta_1$ and $\zeta_2$, we get the relation
\begin{equation}
\left\lbrack \frac{\alpha}{\Pm}\p_\zeta b_\phi + u_\phi \right\rbrack_{\zeta_1}^{\zeta_2}
= \frac{3}{2}\int_{\zeta_1}^{\zeta_2}b_r \, \dd \zeta.
\end{equation} 
We can use this relation between $\zeta_B^-$ and $\zeta_B^+$ to connect the passive-field region and the
force-free region. As mentioned above, we neglect the width of the
intermediate region, so that the right-hand side of the equation vanishes.  Then
\begin{equation}
u_\phi(\zeta_B^+) - u_\phi(\zeta_B^-) = \frac{\alpha}{\Pm}\p_\zeta
b_\phi(\zeta_B^-), \label{eq:const_uphi}
\end{equation} 
where we have used the force-free
condition $\p_\zeta b_\phi(\zeta_B^+)=0$. This boundary condition corresponds to a continuous radial electric field across $\zeta_B$.

Finally, two more conditions are needed in order to constrain the
model. These should come from the two components of the equation of
motion. Integrating this equation across the intermediate region is not
straightforward as it contains terms proportional to
$\tilde\rho\beta_0$, which vary under our simple model from infinity in the
passive-field region to zero in the force-free region. To build a simple model, we first make the simplest assumption that the magnetic field does not vary significantly at the transition between the two regions:
\begin{eqnarray}
b_r (\zeta_B^-)&=& b_r (\zeta_B^+)=  b_{r\rms} + D_B \zeta_B \label{eq:const_br} \\
b_\phi (\zeta_B^-)&=& b_\phi (\zeta_B^+)= b_{\phi\rms} \label{eq:const_bphi}
\end{eqnarray}
Although not well motivated theoretically, this assumption might seem intuitively consistent with the assumption of an infinitely thin transition. 
As shown in the next subsections, it is rather well verified for the azimuthal field but not so well for the radial field. Indeed, the profile of the radial component of the magnetic field can show a significant variation around $\zeta_B$, which is not reproduced by the above description. The comparison with numerical calculations described in the next few subsections, however shows that this model is a useful approximate description, which reproduces most relevant effects. A better description of the boundary conditions is developed in Appendix~A for the special case of vanishing viscosity, and heuristically generalised in Appendix~B to non-vanishing viscosity in order to obtain more precise analytical expressions in a two-zone model.

\subsection{Magnetic field diffusion ($b_{r\rms}$ and $D_B$)}
	\label{sec:diffusion}

 \begin{figure*} 
   \centering
   \includegraphics[width= 12cm]{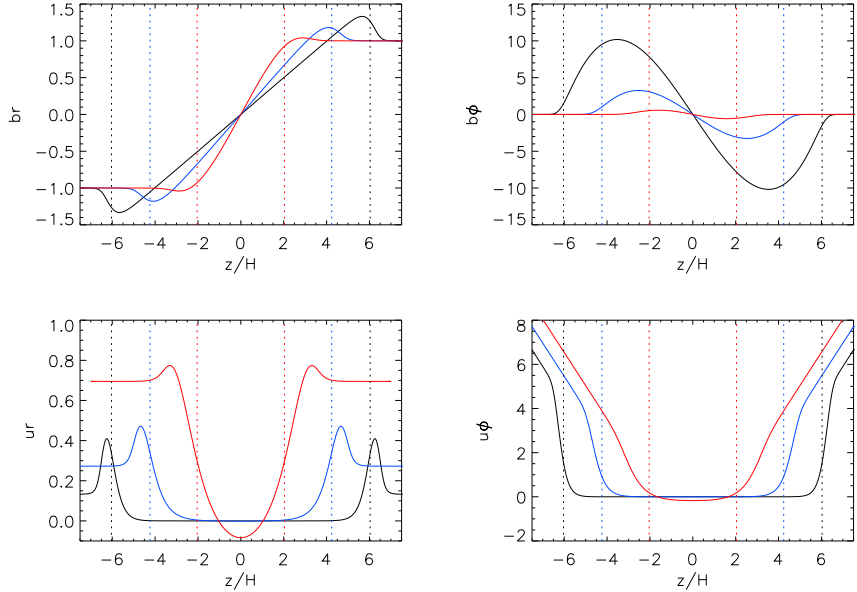}
   \caption{Vertical profiles of the radial (left) and azimuthal
     (right) magnetic field (top) and velocity (bottom) in response to a radial component of the surface
     magnetic field $b_{r\rms}=1$ (i.e.\ the solution vector $\bmath{X}_{br\rms}$). The different colours correspond to different strengths
     of the vertical magnetic field: $\beta_0 = 10$ (red), $\beta_0 =10^4$ (blue), and $\beta_0 =10^8$ (black). The vertical dotted lines illustrate the height $\zeta_B$ at which the magnetic and thermal pressures coincide, marking the transition between the passive and the force-free field regimes.}
   \label{fig:cstalpha_brs0}
 \end{figure*}

 \begin{figure*} 
   \centering
   \includegraphics[width= 12cm]{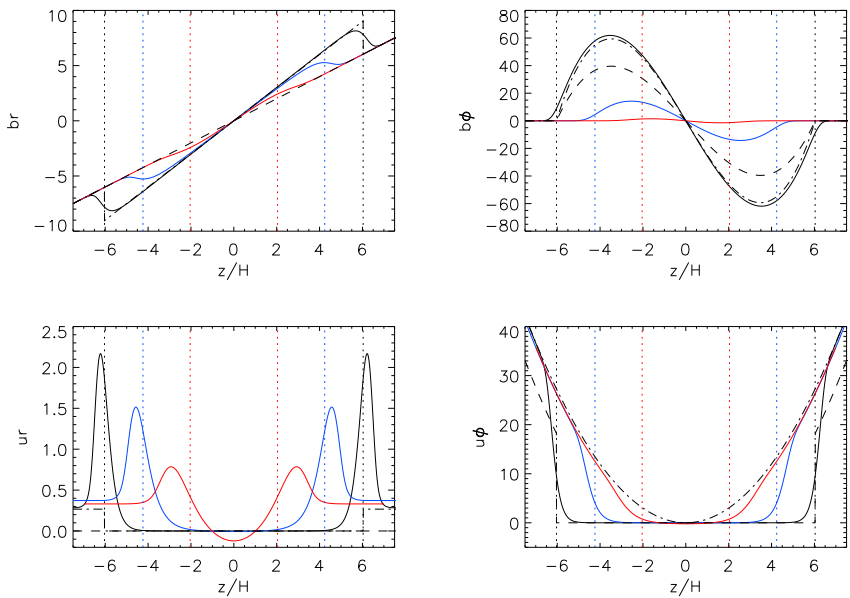}
   \caption{Same as Figure~\ref{fig:cstalpha_brs0} but for the source
     term $D_B=1$ describing the radial gradient of the vertical magnetic field strength (i.e.\ the solution vector $\bmath{X}_{DB}$). The dashed and dot-dashed lines show the prediction of the analytical model  respectively with the simple boundary conditions of Section~\ref{sec:anal} and with the more precise boundary conditions of Appendix~B (only for the case $\beta_0=10^8$).}
   \label{fig:cstalpha_DB0}
 \end{figure*}

We recall that our problem is linear and that the full solution of the problem consists of a superposition of the response of the system to various source terms (see Section~\ref{sec:parameters}).  We consider first the source terms $b_{r\rms}$ and $D_B$, which cause the magnetic field to bend and diffuse through the disc. The profiles computed numerically (including only the source terms $b_{r\rms}$ and $D_B$) are illustrated in Figures~\ref{fig:cstalpha_brs0} and \ref{fig:cstalpha_DB0} for three different strengths of the magnetic field. The quantities that are plotted are the dimensionless magnetic field and velocity variables as defined in Section~\ref{sec:system}.  Note that in each case ($b_{r\rms}=1$ and $D_B=1$) the poloidal magnetic field bends outwards. 

The transition between the passive-field region close to the midplane and the force-free region at $\zeta>\zeta_B$ is clearly visible for the weak fields ($\beta_0 =10^4$ and $10^8$) as the height above which the velocity departs from zero and the magnetic field tends to the values dictated by the boundary conditions at infinity. Note that in the case of $\beta_0=10$ the magnetic field is sufficiently strong so that it does not really behave as a passive field at the midplane. As expected from Section~\ref{sec:passive_field}, in the passive-field region, the velocity vanishes while the radial component of the magnetic field has a linear profile and the azimuthal component looks like a third-order polynomial. In the force-free region, the azimuthal magnetic field vanishes and the radial component is either a constant ($b_{r\rms}$, Figure~\ref{fig:cstalpha_brs0}) or linear ($D_B\zeta$, Figure~\ref{fig:cstalpha_DB0}) as required by the boundary conditions at infinity. Also as expected, the radial velocity is constant and equals the transport velocity of the magnetic flux, while the azimuthal velocity is either linearly increasing or parabolic. 

Let us now use the two-zone model of Section~\ref{sec:anal} to derive approximate vertical profiles. As already mentioned, when only these source terms are included, the velocity vanishes in the passive-field region. The radial magnetic field profile in this region is then linear and can be obtained using equation~(\ref{eq:const_br}):
\begin{equation}
b_r = b_{r1}\zeta = \left(\frac{b_{r\rms}}{\zeta_B} + D_B \right)\zeta.
\end{equation}
Using equation~(\ref{eq:const_bphi}) (with $b_{\phi\rms}=0$), one can obtain the azimuthal magnetic field profile arising from the stretching of this radial field:
\begin{equation}
b_\phi = \frac{1}{4}\frac{\Pm}{\alpha}\left(\frac{b_{r\rms}}{\zeta_B} + D_B \right)\zeta(\zeta^2-\zeta_B^2).
\end{equation}
The transport velocity of the magnetic flux is then obtained from equation~(\ref{eq:const_ur}):
\begin{equation}
 u_{\psi} =  \frac{\alpha}{\Pm} \frac{b_{r\rms}}{\zeta_B},
\label{eq:uflux_mag}
\end{equation}
 and the azimuthal velocity profile in the force-free region is obtained using equation~(\ref{eq:const_uphi}):
\begin{equation}
u_\phi = b_{r\rms}\left(\frac{3}{2}|\zeta| - \zeta_B \right) + \frac{D_B}{4}\left(3\zeta^2-\zeta_B^2 \right).
\end{equation}

At the transition between the two regions, the radial and azimuthal components of the velocity show a jump qualitatively consistent with the boundary conditions of the approximate model (equations~\ref{eq:const_ur}--\ref{eq:const_uphi}). As assumed in the model, the azimuthal component of the magnetic field goes smoothly to zero at $\zeta_B$; however, the radial component has a behaviour that is not described by the simple boundary conditions of Section~\ref{sec:anal}. Instead of smoothly increasing to its final value at infinity, the radial magnetic field reaches a maximum and then decreases to its value at infinity. This can be qualitatively understood with the following argument. At $\zeta\sim\zeta_B$, the azimuthal velocity has to increase and take positive (i.e. super-Keplerian) values, due to the requirement that the radial electric current be continuous (equation~\ref{eq:const_uphi}). This requires a radial Lorentz force directed inward, which corresponds to a radial magnetic field decreasing with $\zeta$. Given that in the passive-field region the radial component of the magnetic field is increasing, it has to show a maximum at the transition around $\zeta\sim\zeta_B$. A quantitative description of this effect is given in Appendix~A and B.

\begin{figure*} 
   \centering
   \includegraphics[width= 12cm]{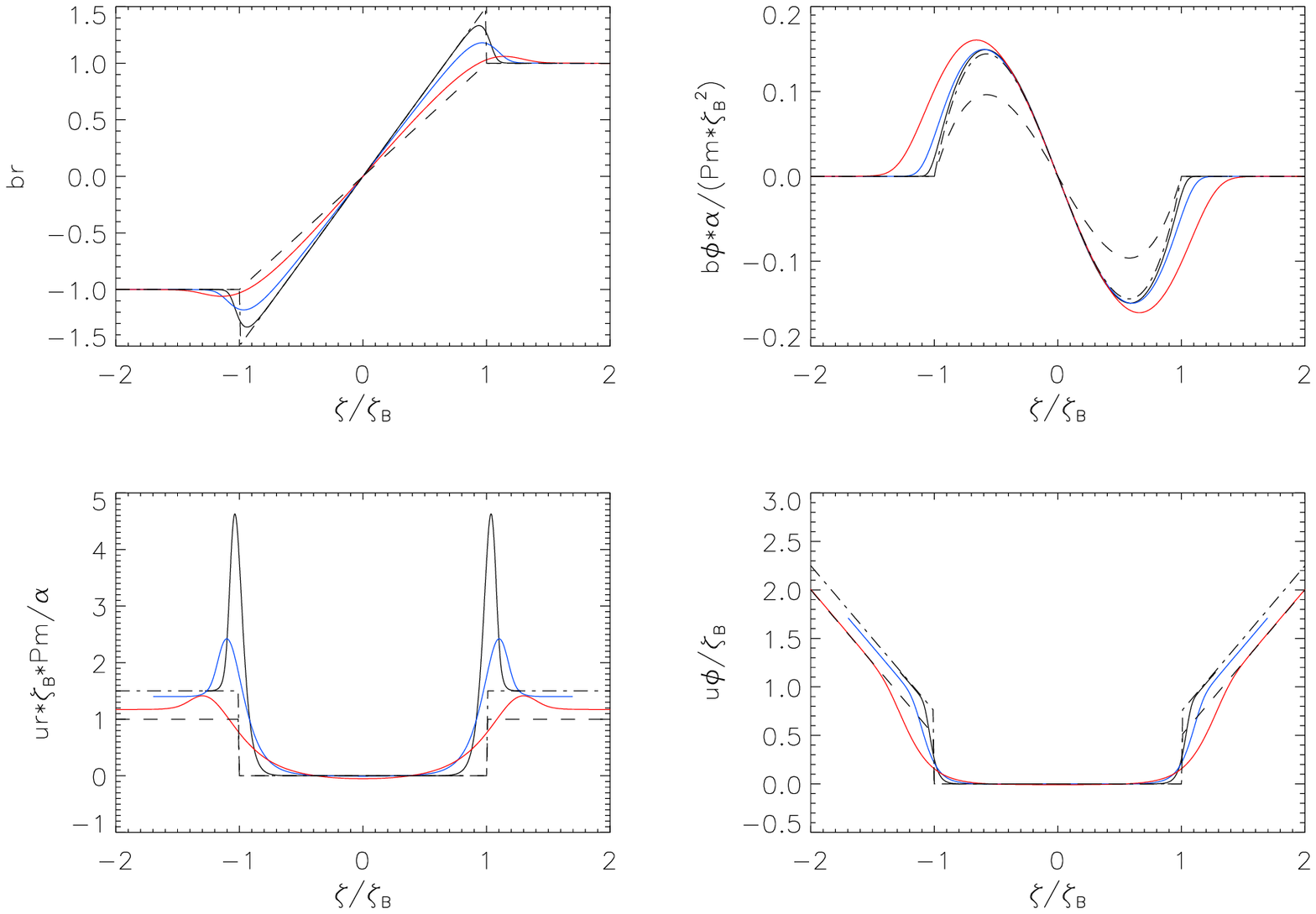}
   \caption{Same as Figure~\ref{fig:cstalpha_brs0} but rescaled
     according to the analytical model of Section~\ref{sec:anal}. The different colours correspond to different strength
     of the vertical magnetic field : $\beta_0 =  10^2$ (red), $\beta_0 =10^4$ (blue), $\beta_0 =10^8$ (black). The dashed and dot-dashed lines show the prediction of the analytical model respectively with the simple boundary conditions of Section~\ref{sec:anal} and with the more precise boundary conditions of Appendix~B. }
   \label{fig:cstalpha_brs_scaled}
 \end{figure*}

\begin{figure*} 
   \centering
   \includegraphics[width= 12cm]{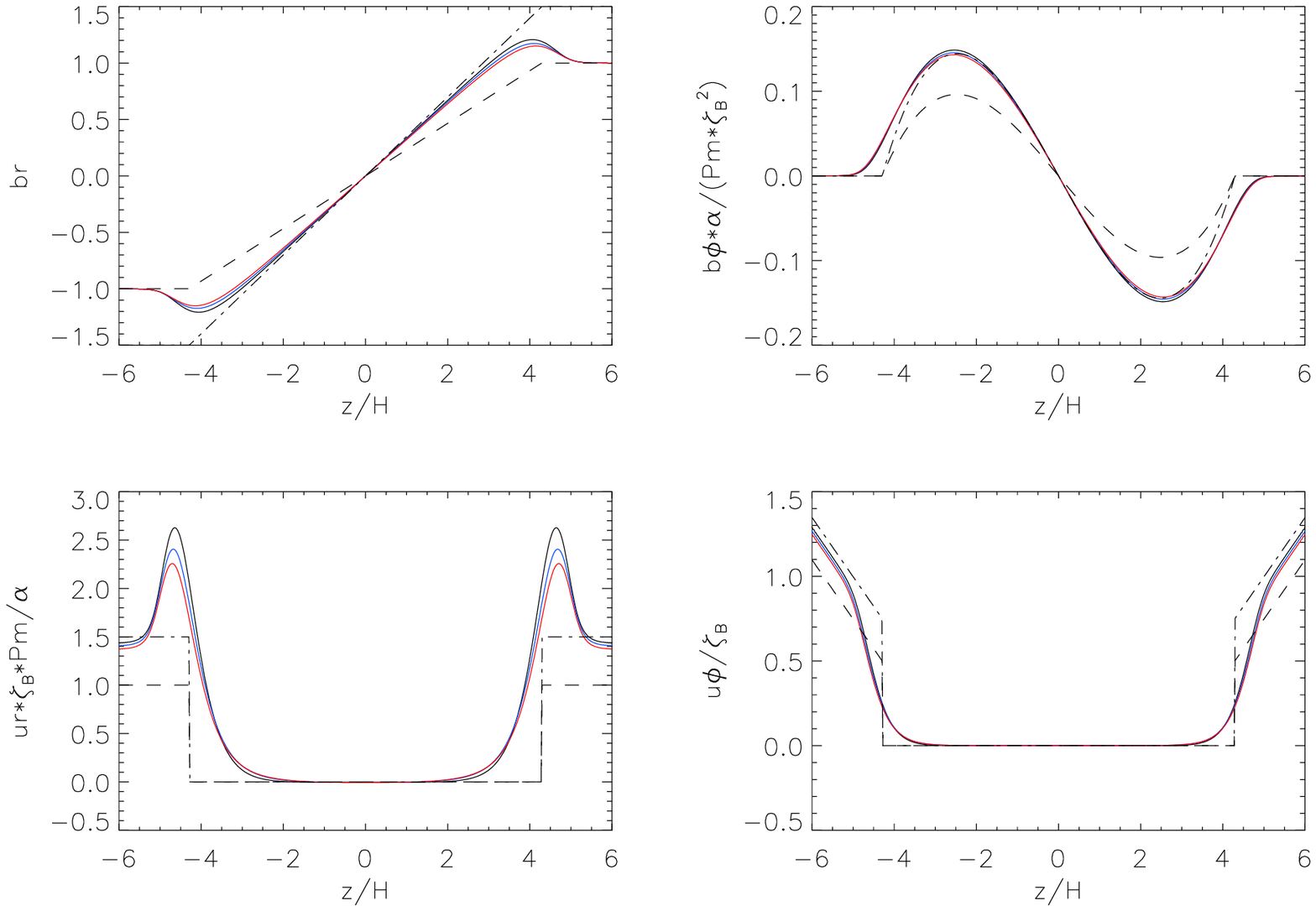}
   \caption{Same as Figure~\ref{fig:cstalpha_brs_scaled} but varying the magnetic Prandtl number: $\Pm =
     0.2$ (black), $\Pm =1$ (blue), and $\Pm =5$ (red).}
   \label{fig:cstalpha_brs_scaled_pm}
 \end{figure*}

To go further in the comparison between the approximate analytical model and the numerically computed
profiles, we remark that the dependence on the magnetic Prandtl number and the magnetic field strength can be scaled out of the analytical expressions in the following way (here for the source term $b_{r\rms}$ only):
\begin{eqnarray}
b_r(\zeta<\zeta_B) &=& b_{r\rms}\frac{\zeta}{\zeta_B}, \label{eq:scaled_br} \\
\frac{\alpha}{\Pm \zeta_B^2} b_\phi(\zeta < \zeta_B) &=&
\frac{b_{r\rms}}{4}\frac{\zeta}{\zeta_B}\left(\frac{\zeta^2}{\zeta_B^2}-1
\right),  \label{eq:scaled_bphi} \\
\frac{\Pm\zeta_B}{\alpha}u_r(\zeta> \zeta_B) &=& b_{r\rms},  \label{eq:scaled_ur} \\
\frac{1}{\zeta_B}u_\phi(\zeta>\zeta_B) &=&
b_{r\rms}\left(\frac{3}{2}\frac{|\zeta|}{\zeta_B}-1 \right).  \label{eq:scaled_uphi}
\end{eqnarray}
In Figures~\ref{fig:cstalpha_brs_scaled} and
\ref{fig:cstalpha_brs_scaled_pm} we show the scaled profiles (i.e. of
$b_r$, $\alpha b_\phi/(\Pm \zeta_B^2)$, $\Pm\zeta_B u_r/\alpha$, and
$u_\phi/\zeta_B$) as a function of $\zeta/\zeta_B$ for different
values of the magnetic field strength (Figure~\ref{fig:cstalpha_brs_scaled}) and of the magnetic Prandtl
number (Figure~\ref{fig:cstalpha_brs_scaled_pm}). The model
suggests that all these curves should lie on top of each other, and
coincide with the dashed curves that represent the analytical profiles
given by equations~(\ref{eq:scaled_br})--(\ref{eq:scaled_uphi}). The different profiles do indeed lie very close to each other (except maybe for the $\beta_0=100$ case, where the passive-field hypothesis at the midplane is less well verified) showing that the dependences on the magnetic field strength and the magnetic Prandtl number have been properly scaled out. 

One exception to this agreement, however, is the bump in radial
velocity close to  $\zeta=\zeta_B$, which is more and more pronounced
as $\beta_0$ increases. The two-zone model fails to reproduce this
feature as it takes place in the intermediate region where the magnetic
field is neither negligible nor dominant. Furthermore, note that the agreement between the numerical profiles and the analytical ones is
not perfect. All the differences can actually be traced back to the profile
of radial magnetic field, which goes through a maximum at $\zeta\sim\zeta_B$ as discussed earlier. The presence of the maximum leads to a larger radial field at $\zeta=\zeta_B^-$ than assumed in the model. The bump in the radial velocity is due to the presence of the maximum in the radial magnetic field, together with the requirement that the azimuthal electric field is uniform. The improved boundary conditions described in the Appendix~B yield a much better agreement between the numerical profiles and analytical ones (shown with dot-dashed lines in Figures~\ref{fig:cstalpha_DB0}, \ref{fig:cstalpha_brs_scaled} and \ref{fig:cstalpha_brs_scaled_pm}).


 \begin{figure*} 
   \centering
   \includegraphics[width= 2\columnwidth]{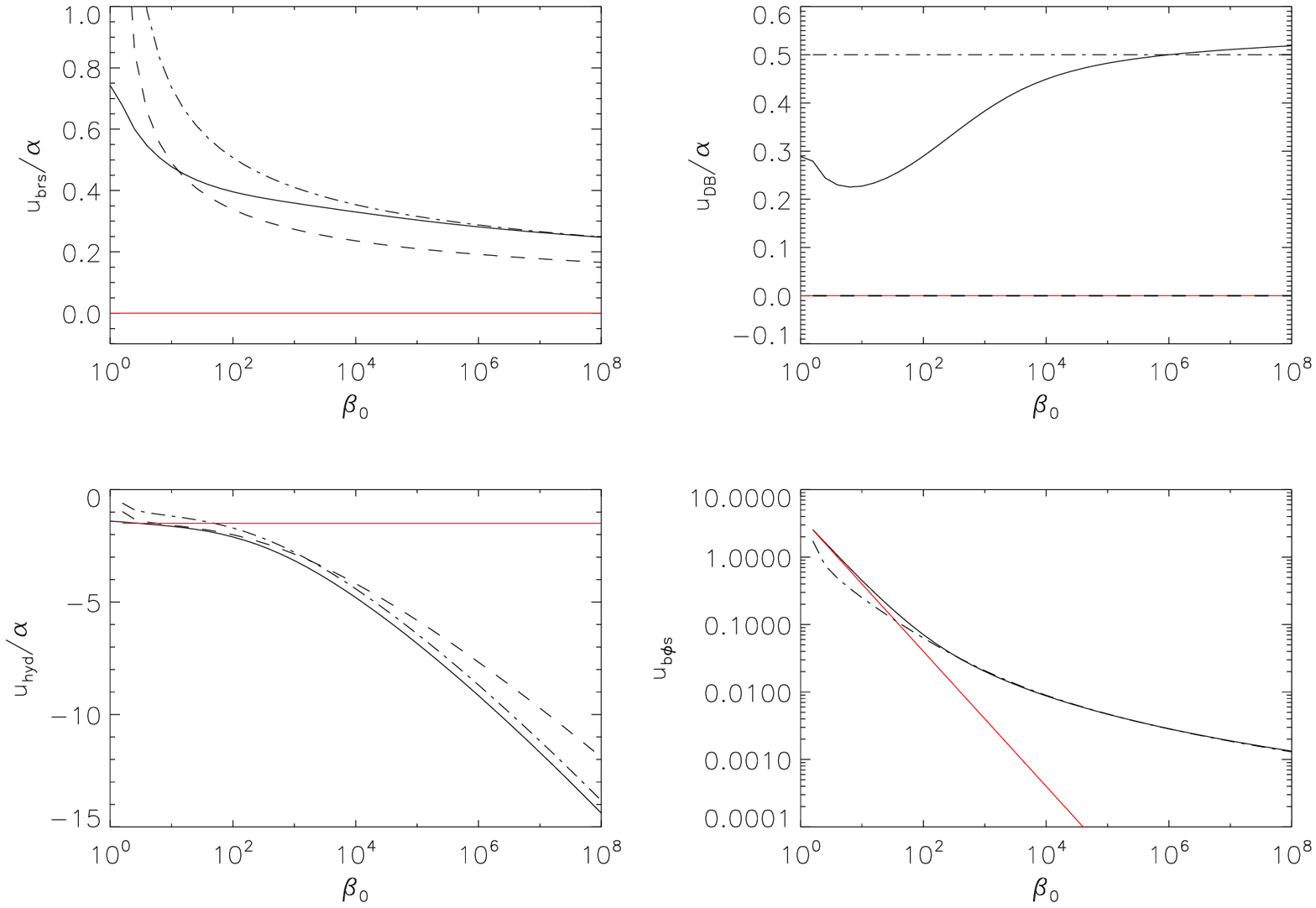}
   \caption{Contributions to the transport velocity of the magnetic flux, $u_\psi$, as a function of the magnetic field strength, for different source terms: diffusion due to the bending across the disc ($b_{r\rms}$ source term, upper left panel), diffusion due to the radial
     gradient of $B_z$ ($D_B$, upper right panel), advection due to the turbulent viscosity (`hydrodynamic' source terms, $D_H=1$ and $D_{\nu\Sigma}=0$, lower left panel), advection due to angular momentum removal by an outflow ($b_{\phi\rms}$, lower right panel). In all panels, the full black line corresponds to the numerical results, while dashed and dash-dotted lines correspond to the analytical model respectively with the simple boundary conditions of Section~\ref{sec:anal} and with the more precise boundary conditions of Appendix~B. The red lines show the corresponding contributions to the mass transport velocity $u_m$.} 
   \label{fig:cstalpha_uflux_beta}
 \end{figure*}

Figure~\ref{fig:cstalpha_uflux_beta} shows the dependence of the transport velocity of the magnetic flux on the strength of the vertical field (upper left and right panels for $b_{r\rms}$ and $D_B$ respectively). The prediction of the analytical model agrees well with the numerical calculation for $\beta_0 > 10^3-10^4$, especially when using the improved boundary conditions of Appendix~B. We recall that for a thin disc and a significant inclination of the surface magnetic field, the dominant contribution to the diffusion is that of the radial field $u_{br\rms}$ (which is larger than $u_{DB}$ by a factor $\sim r/H$). It is interesting to note that this diffusion process is less efficient for weak magnetic fields. Indeed for the parameters explored here, the diffusion is up to a factor of 4 slower than the crude estimate stated in the introduction (equation~(\ref{eq:vdiff}), which corresponds to $u_{br\rms}=1$ in Figure~\ref{fig:cstalpha_uflux_beta}). This slower diffusion is a consequence of the fact that field lines bend on a length-scale $\zeta_B$, which is larger for weak fields.

\subsection{Transport due to angular momentum loss in an outflow ($b_{\phi\rms}$)}
	\label{sec:outflow}
	
 \begin{figure*} 
   \centering
   \includegraphics[width= 12cm]{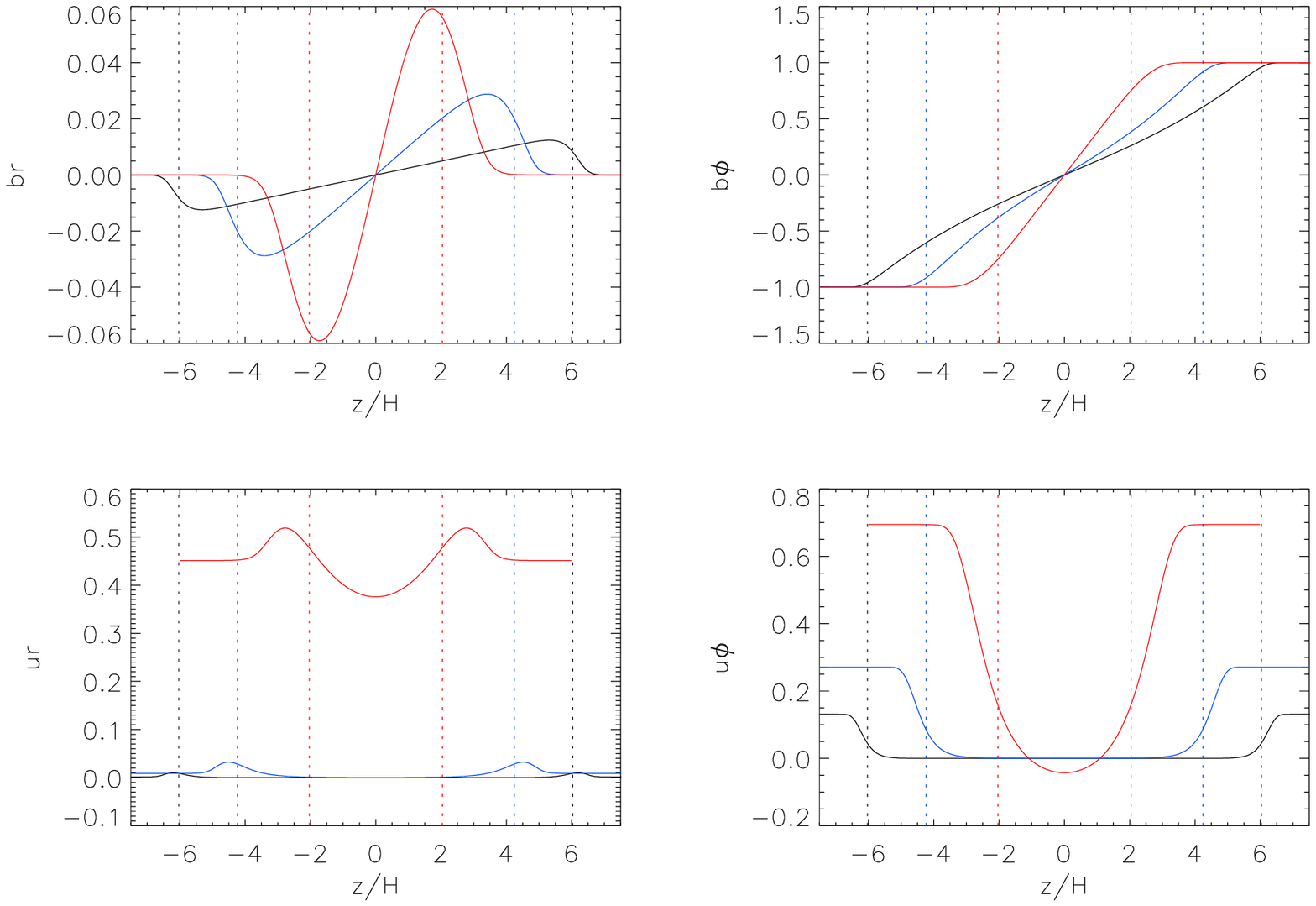}
   \caption{Same as Figure~\ref{fig:cstalpha_brs0} but for an azimuthal surface magnetic field $b_{\phi\rms}=1$. Note that, in reality, $b_\phi\rms$ would be expected to be negative (so that the outflow removes angular momentum from the disc), so this component of the solution would be multiplied by a negative coefficient.}
   \label{fig:cstalpha_bphis}
 \end{figure*}

Applying the two-zone model with the simple boundary conditions of Section~\ref{sec:anal} predicts that no advection is induced by the source term $b_{\phi\rms}$. Indeed, in the passive-field region the velocity vanishes, as does the radial magnetic field. Alone, the azimuthal magnetic field has a linear dependence on $\zeta$: 
\begin{equation}
b_\phi = b_{\phi\rms}\frac{\zeta}{\zeta_B}
	\label{eq:bphis_bphi}
\end{equation}
The transport of magnetic flux vanishes: $u_\psi=0$, while the azimuthal velocity jumps around $\zeta_B$ to its final value:
\begin{equation}
u_\phi = \frac{\alpha}{\Pm}\frac{b_{\phi\rms}}{\zeta_B}
	\label{eq:bphis_uphi}
\end{equation}

This is not the whole story however, because two effects have not been taken into account. First, the jump in radial magnetic field around $\zeta_B$ (Figure~\ref{fig:cstalpha_bphis}) induces some transport of magnetic flux. This is described quantitatively in Appendix~B, and the analytical prediction for the transport velocity of magnetic flux is compared to numerical results in Figure~\ref{fig:cstalpha_uflux_beta}. The agreement is very good for rather weak fields ($\beta_0>10^2$). For larger values of the magnetic field, another effect neglected in the two-zone model becomes important. Indeed, the magnetic force cannot be neglected anymore and the torque exerted by the azimuthal field induces a non-negligible radial velocity, which contributes to the transport of magnetic flux. For $\beta_0<100$, the transport velocity of magnetic flux is then very close to that of mass (given by equation~(\ref{um})).

As might have been expected, the efficiency of the surface azimuthal magnetic field to transport mass and magnetic flux strongly increases as the magnetic field strength is increased (Figure~\ref{fig:cstalpha_uflux_beta}). It is interesting to note however that the transport velocity of magnetic flux is much larger than that of mass, when the magnetic field is rather weak.

\subsection{Advection due to the turbulent viscosity}
	\label{sec:hydro}
 \begin{figure*} 
   \centering
   \includegraphics[width= 12cm]{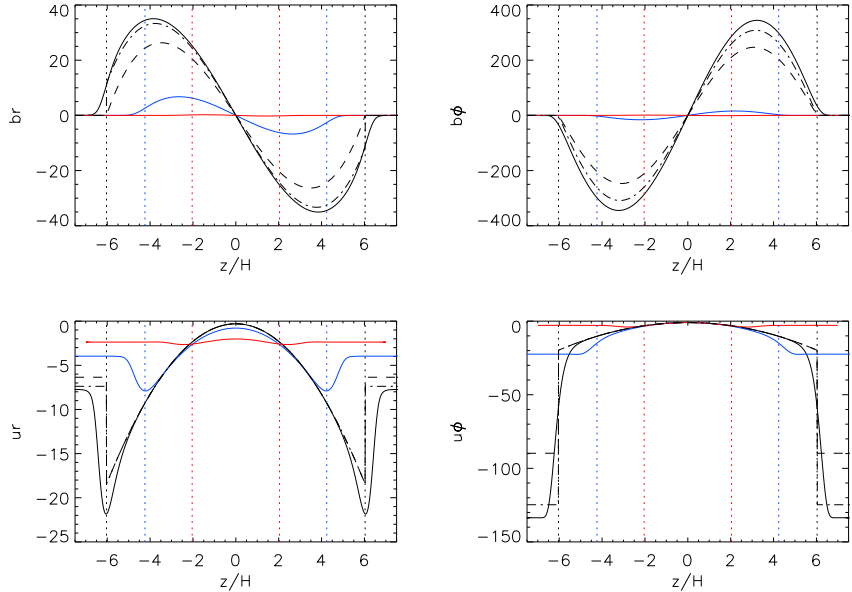}
   \caption{Same as Figure~\ref{fig:cstalpha_DB0} but for the
    `hydrodynamic' source terms $D_H=1$ and $D_{\nu\Sigma}=0$.}
   \label{fig:cstalpha_hydro0}
 \end{figure*}

All the remaining source terms ($D_{\nu\Sigma}$, $D_H$ and those which solution vector is $\mathbf{X}_K$) are hydrodynamical in nature and describe the advection process due the turbulent viscosity. First, we note that the source term $D_{\nu\Sigma}$ admits an exact solution with uniform radial and azimuthal velocities and vanishing radial
and azimuthal magnetic fields:
\begin{eqnarray}
b_r &=&0, \\
b_\phi &=&0, \\
u_r &=&-3\alpha D_{\nu\Sigma}, \\
u_\phi &=& \frac{D_{\nu\Sigma}}{2}.
\end{eqnarray}

We now focus on the other hydrodynamical source terms, and their solution in the approximate analytical model. In the passive-field region the velocity profiles are the parabolae given by equations~(\ref{eq:ur_hydro})--(\ref{eq:uphi2}) (note that the parabolic shape is clear in Figure~\ref{fig:cstalpha_hydro0}). The magnetic field profiles result from the stretching of the magnetic field lines by the parabolic velocity profile, and from the condition that both radial and azimuthal components vanish at $\zeta=\zeta_B$ (for the simple boundary conditions of Section~\ref{sec:anal}):
\begin{equation}
b_r = \frac{\Pm}{\alpha}\frac{u_{r2}}{3}\zeta(\zeta_B^2-\zeta^2), 
\end{equation}
\begin{equation}
b_\phi = \frac{\Pm}{\alpha}(\zeta_B^2-\zeta^2)\left\lbrack\frac{u_{\phi2}}{3}\zeta +
  \frac{u_{r2}}{120}\frac{\Pm}{\alpha}\zeta(3\zeta^2-7\zeta_B^2) \right\rbrack. 
\end{equation}
Then the velocities in the force-free region and, most importantly, the advection speed of the magnetic flux are found using equations~(\ref{eq:const_ur}) and~(\ref{eq:const_uphi}):
\begin{equation}
 u_{\psi} = u_{r0} + \frac{u_{r2}}{3}\zeta_B^2,  \label{eq:uflux_hydro} 
\end{equation}
\begin{equation}
u_\phi(\zeta>\zeta_B) = u_{\phi0} + \frac{u_{\phi2}}{3}\zeta_B^2 + \frac{u_{r2}}{15}\frac{\Pm}{\alpha}\zeta_B^4.
\end{equation}
These profiles are compared with the numerical solution in Figure~\ref{fig:cstalpha_hydro0} for the case $\beta_0=10^8$ (the
analytical profile is the dashed line). The agreement is acceptable though the magnetic profiles are slightly underestimated. Again, a quantitatively better agreement is found when using the improved boundary conditions of Appendix~B (dot-dashed lines).

Rewriting the transport velocity of the magnetic flux as a function of the source terms, one finds
\begin{equation}
\frac{u_{\psi}}{\alpha} = -\frac{9}{2} + 5D_H - 3D_{\nu\Sigma} +
\frac{3-5D_H}{4\alpha^2+1}\left(4\alpha^2 + \frac{\zeta_B^2}{3}
\right).  \label{eq:uflux_hydro_bis}
\end{equation}
This shows that the inward transport is faster for weak fields (i.e.\ large $\zeta_B$) if $D_H>0.6$. It is for example the case for our fiducial values of $D_H=1$ and $D_{\nu\Sigma}=0$:
\begin{equation}
u_{\psi,\mathrm{hyd}} = \alpha\left\lbrack \frac{1}{2} - \frac{2}{4\alpha^2+1}\left(4\alpha^2+\frac{\zeta_B^2}{3} \right) \right\rbrack.
	\label{eq:uflux_hyd}
\end{equation}
The lower left panel of Figure~\ref{fig:cstalpha_uflux_beta} compares the transport velocity of the magnetic flux $u_\mathrm{hyd}$ (full black line) with equation~(\ref{eq:uflux_hyd}) (dashed line) and the transport velocity of mass (red line). Interestingly, the transport velocities of mass and magnetic flux coincide for strong fields $\beta_0<10$, but for weak fields the advection of magnetic flux is much faster than that of mass (by a factor up to 10 for the present parameters). This behaviour is correctly reproduced by the approximate analytical model, and its interpretation will be discussed in the following subsection.

\subsection{Advection/diffusion speed as a vertical average}
	\label{sec:uflux_average}

The results of Sections~\ref{sec:diffusion}, \ref{sec:outflow} and \ref{sec:hydro} show that the transport of magnetic flux is significantly affected by the vertical structure, in the sense that it differs from the simple vertical average used by many previous works such as \citet{lubow94a}. This can be understood by noting that the transport velocity of magnetic flux, contrary to that of mass, is not given by a density weighted average. Below, we repeat with our notation the argument given by \citet{ogilvie01} on the correct way of averaging the equations in order to obtain the effective transport velocities of mass and magnetic flux. We then use it to interpret our results.

 Dividing equation~(\ref{upsi}) by the dimensionless resistivity $\alpha/\Pm$,
and integrating between $\zeta=0$ and
$\zeta$, we obtain:
\begin{equation}
 u_{\psi} = \bar{\eta}\left(\frac{b_r}{\zeta} - D_B \right) + \bar{u}_r,
 	\label{eq:upsi_average}
\end{equation}
where $\bar{\eta}$ is an average dimensionless resistivity defined in the following way (it is
actually the inverse of the height-averaged conductivity):
\begin{equation}
\bar{\eta} = \frac{\zeta}{\int_0^\zeta \frac{\Pm}{\alpha}\, \dd
  \zeta^{\prime}},   \label{eq:eta_average}
\end{equation}
and $\bar{u}_r$ is the average of the radial velocity weighted by the conductivity ($1/\eta$):
\begin{equation}
\bar{u}_r = \frac{1}{\int_0^\zeta \frac{\Pm}{\alpha}\, \dd \zeta^{\prime}}
\int_0^\zeta \frac{\Pm}{\alpha} u_r\, \dd \zeta^{\prime}.   \label{eq:vadv_average}
\end{equation}

This averaging procedure shows that the contribution of advection to the velocity at which
the magnetic flux is transported is the vertically averaged radial
velocity weighted by the electrical conductivity.  This average may be
very different from the density-weighted average $u_m$ that describes mass transport. 

Note that equation~(\ref{eq:upsi_average}) is valid for any height $\zeta$ up to which the averaging procedure is performed, and which can be chosen in the most convenient way. Integrating up to an infinite height may not be very useful as the force-free regime would dominate the average and one would simply recover $v_\psi=C_1$, as was already found in Section~\ref{sec:asymptotic}. Instead, the two-zone model developed in Section~\ref{sec:anal} suggests that performing the average up to the height $\zeta_B$ is most useful. By doing so, one can indeed recover the results for $ u_{\psi}$ of Sections~\ref{sec:diffusion} and \ref{sec:hydro} in a simpler way:
\begin{equation}
 u_{\psi} = \frac{\alpha}{\Pm} \left(\frac{b_r(\zeta_B^-)}{\zeta_B} - D_B \right) + \bar{u}_r,
\label{eq:uflux_average_zetaB}
\end{equation}
where $\bar{u}_r$ is the average advection velocity as defined by equation~(\ref{eq:vadv_average}), performing the averaging
procedure between $\zeta=0$ and $\zeta=\zeta_B^-$, and we used $\bar{\eta}=\alpha/\Pm$ since the diffusion coefficients are uniform. Following the two-zone modelling of Section~\ref{sec:anal}, the average advection velocity can be computed using the purely hydrodynamical velocity profile (because the field is assumed to be passive for $\zeta<\zeta_B^-$):
\begin{equation}
\bar{u}_r = \frac{1}{\zeta_B}\int_0^{\zeta_B} (u_{r0}+u_{r2}\zeta^2) \, \dd\zeta =  u_{r0} + \frac{u_{r2}}{3}\zeta_B^2.
\end{equation}
$\bar{u}_r$ gives the value of the advection speed due to the hydrodynamic source terms, as described by equation~(\ref{eq:uflux_hydro}). The value of $b_r$  at $\zeta_B^-$ then has to be prescribed by the boundary conditions. Using that of Section~\ref{sec:anal} gives $b_r(\zeta_B^-) = b_{r\rms}+D_B\zeta_B$, and one recovers equation~(\ref{eq:uflux_mag}). Using the boundary conditions of Appendix~B is also possible in principle but it requires first solving for the full vertical profiles.

The above argument allows for a deeper interpretation of the higher advection speed of weak fields, which was observed in the previous section. Indeed, the mass transport velocity $u_m$ corresponds to a density-weighted average and is therefore dominated by the velocity close to the midplane, where the density is large. By contrast, the magnetic flux transport velocity $u_\psi$ is a conductivity-weighted average, which for the case of uniform diffusion coefficients is equivalent to a height-average. As a consequence the advection of a weak field is significantly affected by the low-density region away from the midplane. This low-density region at high $\zeta$ can have a much faster inward radial velocity than in the midplane, if $u_{r2}<0$ as is readily found for standard parameters. Such parabolic velocity profiles, where the radial velocity becomes more and more negative with the distance from the midplane, are well known solutions of viscous hydrodynamical disc models \citep[e.g.][]{takeuchi02}.

\subsection{Inclination of the field lines}
	\label{sec:inclination}
	
 \begin{figure*} 
   \centering
   \includegraphics[width= 12cm]{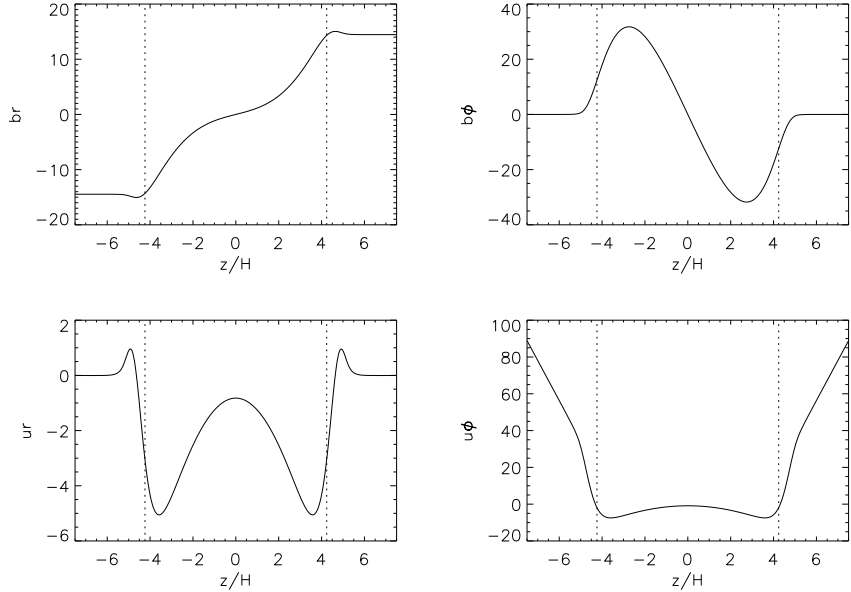}
   \caption{Profiles of the radial (left) and azimuthal (right) components of the magnetic field (upper) and the velocity (lower), in a stationary situation where the advection of magnetic flux is exactly compensated by the diffusion of the radial field across the disc. The parameters used are: $\beta_0=10^4$, `hydrodynamic' source terms $D_H=1$ and $D_{\nu\Sigma}=0$, and a radial surface magnetic field $b_{r\rms}\simeq14.5$ chosen so that diffusion compensates advection (note, however, that our model neglects the magnetic compression of the disc).}
   \label{fig:cstalpha_stat}
 \end{figure*}
 
  \begin{figure} 
   \centering
   \includegraphics[width=\columnwidth]{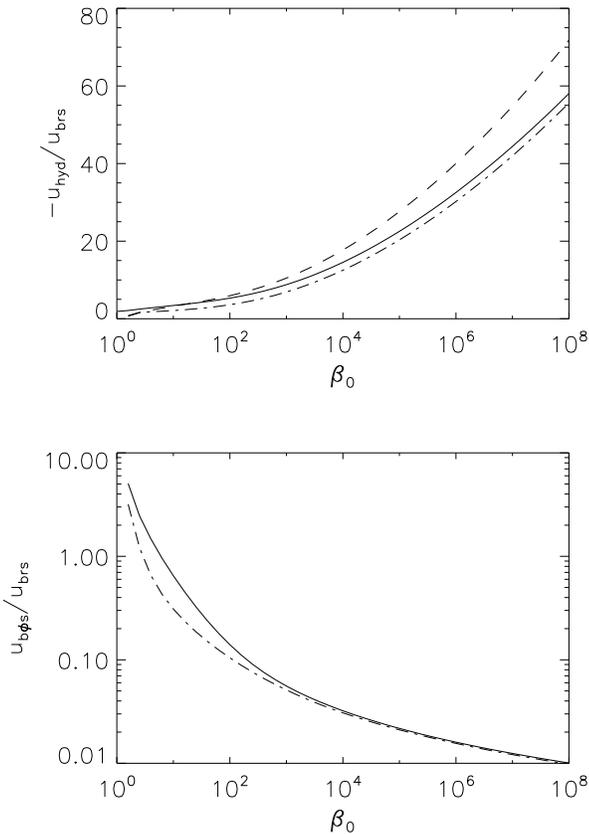}
   \caption{Radial inclination of the field lines at the surface of the disc obtained in a stationary situation. The upper panel shows the inclination ($B_{r\rms}/B_z$) in units of $H/r$ such that the diffusion compensates the advection due to the turbulent viscosity (`hydrodynamic' source terms). The lower panel shows the ratio $-b_{r\rms}/b_{\phi\rms}$ such that diffusion compensates advection due to an outflow. In both panels, the full black line corresponds to the numerical results, while dashed and dash-dotted lines correspond to the analytical model respectively with the simple continuous magnetic field boundary conditions (Section~\ref{sec:anal}) and with the more precise boundary conditions of Appendix~B. Note that in the lower panel the dashed line is not shown as it would equal 0.}
   \label{fig:inclination}
 \end{figure}

As outlined in the introduction, one of the most important diagnostics of the relative efficiency of the magnetic flux advection and diffusion processes is the inclination of the field lines at the surface of the disc that can be achieved in a stationary situation. In such a stationary situation the diffusion velocity due to the inclination, $u_{br\rms}b_{r\rms}$ (recall that $b_{r\rms}=r/H\times B_{r\rms}/B_z$), compensates the advection velocity $u_\mathrm{hyd}$ (as an example, let us focus on the case $D_H=1$, $D_{\nu\Sigma}=0$). Figure~\ref{fig:cstalpha_stat} illustrates the vertical profiles obtained in such a stationary situation. The inclination is then:
\begin{equation}
\frac{B_{r\rms}}{B_z} = -\frac{H}{r}\frac{u_\mathrm{hyd}}{u_{br\rms}}.
\end{equation}
In the limit of an infinitely thin disc, and if $u_\mathrm{hyd}$ and $u_{br\rms}$ are of order unity, one recovers the classical result that the inclination is vanishingly small. The inclination of the field lines can be estimated with the approximate analytical model (here with the simple continuous magnetic field boundary condition of Section~\ref{sec:anal}):
\begin{equation}
\frac{B_{rs}}{B_z} = \frac{H}{r}\Pm\zeta_B\left\lbrack  -\frac{1}{2} +
  \frac{2}{4\alpha^2+1}\left(4\alpha^2+\frac{\zeta_B^2}{3} \right) \right\rbrack.
\end{equation}
In Figure~\ref{fig:inclination}, this estimate (represented by dashed lines) is compared to the numerical result. The more precise estimate represented in dash-dotted line is given in Appendix~B. As expected, the formula shows that the bending does not depend much on the value of $\alpha$ (at least if $\alpha^2\ll1$) but rather on the magnetic Prandtl number $\Pm$. In particular one recovers the result that a significant inclination can be achieved if $\Pm \sim r/H$ (but this is not expected to be true). The formula and associated figure however suggest a new way to achieve a large inclination: to be in the weak field regime where $\beta_0$ and $\zeta_B$ are large. Indeed, a significant inclination is possible for weak fields if the disc is not too thin. For example, if $H/r=0.1$ a field of $\beta_0\sim10^3$ can be significantly bent ($B_r\sim B_z$). This is due both to a reduced diffusion and a faster advection speed as compared to the rough estimates quoted in the introduction (as discussed in previous subsections).

Similarly the ratio of the radial and azimuthal components of the surface magnetic field can be estimated in a stationary situation where diffusion is compensated by the advection due to an outflow:
\begin{equation}
\frac{B_{r\rms}}{B_{\phi\rms}} = - \frac{u_{b\phi\rms}}{u_{br\rms}}.
	\label{eq:inc_bphis}
\end{equation}
(Recall that $B_{\phi\rms}$ is negative so that the outflow removes angular momentum from the disc). This ratio is represented in the lower panel of Figure~\ref{fig:inclination} and compared with the estimate of the analytical model given in Appendix~B. As might be expected, the advection due the outflow is more efficient at bending the magnetic field for strong magnetic fields. 

In order to relate the above ratio to a radial inclination of the field with respect to the vertical, one would need to know the value of the azimuthal magnetic field at the surface, which determines the efficiency with which the outflow removes angular momentum. This would require a study of the launching and acceleration of the outflow, which is beyond the scope of this paper. We therefore keep it as a free parameter and obtain the following expression for the inclination: 
\begin{equation}
\frac{B_{r\rms}}{B_z} = -\frac{B_{\phi\rms}}{B_{z}} \frac{u_{b\phi\rms}}{u_{br\rms}}.
\end{equation}
For a given value of $B_{\phi\rms}/B_{z}$, Figure~\ref{fig:inclination} suggests that only magnetic fields above a certain strength can sustain a significant inclination of the field lines, if the advection is due to an outflow. For $B_{\phi\rms}/B_{z} = -1$, the threshold is $\beta_0\sim10$, while for $B_{\phi\rms}/B_{z} = -10$ it reaches $\beta_0\sim10^3$. Although magneto-centrifugal acceleration is usually thought to produce an outflow if the magnetic field is strong ($\beta_0$ of order or slightly above unity), we note that outflows with weaker fields may be driven by discs with superheated surface layers. Indeed, \citet{murphy10} have recently reported the production of a magneto-centrifugal outflow from a weakly magnetised disc ($\beta_0 \sim 100-1000$). This opens the possibility that an outflow from a weakly magnetised disc could significantly affect the transport of magnetic flux.

\subsection{A scenario to obtain strong inclined magnetic fields}
	\label{sec:scenario}

\begin{figure} 
   \centering
   \includegraphics[width=\columnwidth]{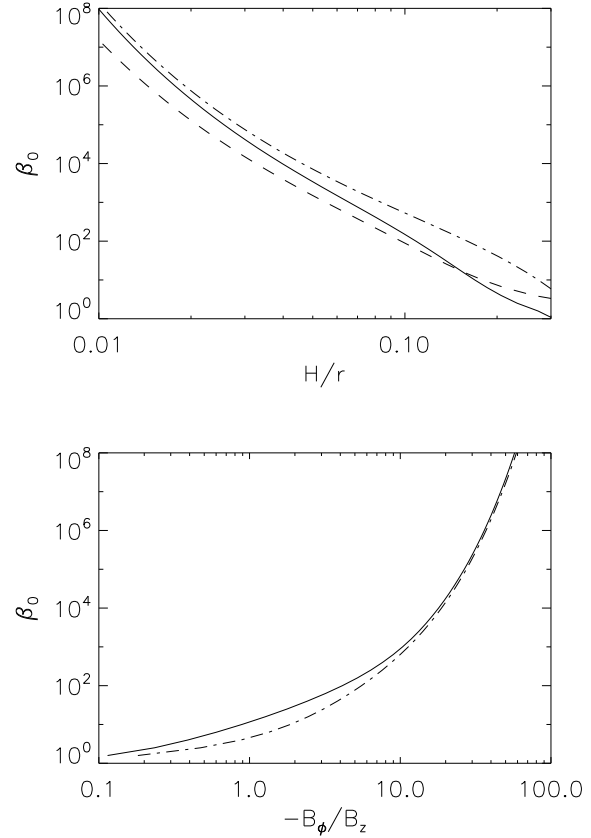}
   \caption{Upper panel: maximum strength of the magnetic field (i.e. minimum $\beta_0$) for which an inclination of $30\degree$ of the surface magnetic field can be sustained, if the advection is due to the turbulent viscosity. This maximum strength is plotted as a function of the aspect ratio of the disc $H/r$. Bottom panel: minimum strength of the magnetic field (i.e. maximum $\beta_0$) for which an inclination of $30\degree$ of the surface magnetic field can be sustained, if the magnetic flux advection is due to an outflow. This minimum strength is shown as a function of the azimuthal magnetic field component at the surface of the disc, parametrising the efficiency at which the outflow extracts angular momentum from the disc.}
   \label{fig:beta_stationary}
 \end{figure}

\begin{figure} 
   \centering
   \includegraphics[width=\columnwidth]{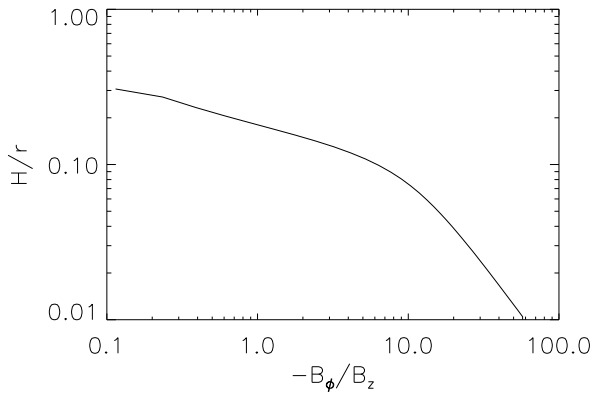}
   \caption{Aspect ratio above which a disc can accumulate magnetic flux and create strong magnetic fields in the scenario outlined in Section~\ref{sec:scenario}. This aspect ratio depends strongly on the efficiency at which an outflow removes angular momentum, parameterised by $-B_{\phi \rms}/B_z$.}
   \label{fig:aspect_ratio}
 \end{figure}

The above discussion suggests a scenario by which strong and inclined magnetic fields could be obtained in a moderately thin disc through the inward advection of an initially weak magnetic field. In such a disc, the advection of a weak field by the accretion flow due to the turbulent viscosity is possible up to a certain strength, which depends strongly on the aspect ratio of the disc (Figure~\ref{fig:beta_stationary}, upper panel). Unless the disc is extremely thick, this strength is however rather weak: for example $\beta_0\sim10^3$ is reached for an already rather thick disc with $H/r > 0.07$. Creating and sustaining a stronger inclined field might then be possible if the advection of the magnetic flux is primarily due to an outflow\footnote{Note that this does not mean that the angular momentum transport and hence the advection of mass is primarily due to the outflow, since the advection speed of mass and magnetic flux can be very different.}, which is possible above a certain strength of the magnetic field. The minimum strength allowing the outflow to compensate diffusion is shown as a function of $B_{\phi\rms}/B_z$ (parameterising the ability of an outflow to remove angular momentum from the disc) in Figure~\ref{fig:beta_stationary} (lower panel). 

A condition for this scenario to work is that the maximum strength for advection through turbulence should be larger than the minimum strength for advection through an outflow. This requirement is fulfilled above a certain aspect ratio, shown as a function of $B_{\phi\rms}/B_z$ in Figure~\ref{fig:aspect_ratio}. This suggests that for reasonable values of the surface azimuthal field, the disc needs to be rather thick to enable the production of a strong magnetic field: $H/r > 0.2$ for $B_{\phi\rms}/B_{z} = -1$, or $H/r>0.07$ for $B_{\phi\rms}/B_{z} = -10$. 

Obviously, these numerical values should be taken with a grain of salt for several reasons. The effect of an outflow has been taken into account in a quite approximate way. A proper description of this scenario would need to include the launching and acceleration process of the jet, which should determine self consistently the value of the azimuthal component of the magnetic field at the surface of the disc. Furthermore a variation of the diffusion coefficients in the vertical direction might be expected to change quantitatively the result, as will be studied in paper~II. The turbulence may also not have such a simple effect as isotropic diffusion coefficients ; for example, the vertical diffusion may be expected to be less efficient than the radial one due to smaller vertical velocities in MRI turbulence. As a consequence, the numerical values for the minimum aspect ratio should not be considered as a strong constraint, and it might turn out that this scenario is effective for somewhat lower aspect ratio.

\section{Conclusion and discussion}
	\label{sec:discussion}

We have developed a formalism enabling us to compute self-consistently the radial transport of poloidal magnetic flux and mass in a thin accretion disc, where the effect of turbulence is modelled by an effective viscosity and resistivity. For this purpose, we performed a systematic asymptotic expansion in the limit of a thin disc.  We assumed that the magnetic field is only weakly inclined with respect to the vertical direction, in order to avoid the complication of modelling a magneto-centrifugal outflow, as well as for self-consistency in the expansion. Owing to this assumption, the solution (and thus the transport velocity of the magnetic flux) depends linearly on a number of source terms. These source terms are the radial gradients of various quantities (surface density, aspect ratio, magnetic field strength), and non-vanishing values of the radial and azimuthal magnetic field at the surface of the disc. Each of them describes a different physical process which can then be studied and understood independently from the others: the diffusion due the bending of the field lines across the disc ($b_{r\rms}$), the diffusion due the radial gradient of its intensity ($D_B$), the advection due to angular momentum loss in an outflow ($b_{\phi\rms}$), and that due the turbulent viscosity (`hydrodynamic' source terms).

An important aspect of this calculation is that it takes into account the vertical structure of the disc. The vertical profiles of velocity and magnetic field are indeed explicitly computed. The formalism also allows for a vertical dependence of the diffusion coefficients. In this paper, we have however only considered the idealised case of uniform coefficients, leaving the question of their vertical structure for paper II. Even in this simplest case, we found the vertical structure to have an important impact on the transport of magnetic flux. Indeed, the simple vertical averaging usually performed gives results that can be in error by a factor of 10 or more, because of its improper treatment of the vertical structure.  We have developed an approximate analytical model which is able to describe this behaviour, and provides a physical explanation. 
	
Two main effects of the vertical structure on the transport of magnetic flux have been uncovered by this study. Both of these effects are important when the poloidal magnetic field is rather weak in the sense that the magnetic pressure is small compared to the thermal pressure in the midplane. First, we showed that the diffusion of weak fields is less efficient because the bending of the field lines takes place on a larger vertical scale. Second, the advection of magnetic flux can be significantly faster than that of mass, owing to fast radial velocities in the low-density regions away from the midplane. 

Taken together, these two effects enable a more efficient inward transport of magnetic flux, which partly alleviates the long-standing problem of too fast outward diffusion of an inclined magnetic field. Indeed, they suggest that in a moderately thin disc, magnetic fields below a certain strength can be advected inwards without significant outward diffusion. This critical strength depends steeply on the aspect ratio of the disc, however, and is probably rather weak for reasonably thin discs. We propose a scenario in which the advection due to the presence of an outflow could then do the rest of the job, and provide a means to obtain yet stronger magnetic fields. A condition for a successful advection is that the aspect ratio of the disc be larger than a critical value, which is dependent on the efficiency of the outflow to extract angular momentum from the disc. Confirming or invalidating this scenario will require a better understanding of the magnetic flux advection induced by an outflow with a weak magnetic field.

The faster advection of weak magnetic fields could also have interesting consequences on the time-dependence of magnetised accretion discs. Indeed, the magnetic flux distribution would have the counterintuitive ability to evolve significantly faster than the mass distribution. These two potential consequences should be studied in a global time-dependent disc model, where the inclination at the disc surface is computed self-consistently as a result of the distribution of magnetic flux. The time-evolution of the mass and magnetic flux could then be described using the transport rates computed with the formalism described in this paper (with the equations given in Section~\ref{sec:asymptotic}). This question is left for future work.

One limitation of our analysis is that it is restricted to small values of the inclination of the magnetic field with respect to the vertical direction. By comparing with direct numerical simulations of a stratified shearing box, we found that the small-inclination approximation gives good results even at significant inclination as long as an outflow is not launched ($i<30\degree$) and the field is not very strong ($\beta_0 \sim 10^2-10^4$). However, it is expected to break down at stronger fields because magnetic compression would become important, or at still larger inclination, in which case the launching process of an outflow should be included. 

A related worry could come from the large values of $b_\phi$ obtained in some cases for weak magnetic fields (e.g. Figure~\ref{fig:cstalpha_hydro0}). If the aspect ratio is only moderately small (e.g. $H/r \sim 10^{-2}-10^{-1}$) then the azimuthal component of the magnetic field may be larger than the vertical one. In addition to contradicting the initial assumption (in particular the neglect of compression), one might worry about the stability of such a solution.

We have underlined in this paper the importance of the vertical profile of radial velocity. \cite{Fromang11} have studied this profile by means of global MHD simulations and found that it differed from the prediction of viscous alpha models. This comes from the different vertical profile of the turbulent stress, compared to a viscous model with a viscosity independent of height (as was assumed in this paper). This shows the limitation of the idealised model studied in this paper. In paper II, we will consider a vertical profile of the viscosity in order to obtain a more realistic vertical profile of the stress. This study confirms the robustness of our main result: that the magnetic flux is advected faster than mass when the field is weak.

Both the faster advection and the reduced diffusion of weak magnetic fields rely on the dynamics of the low-density region away from the midplane. Turbulent MHD simulations both of stratified shearing boxes and global models show that this region tends to be strongly magnetised \citep{miller00,fromang06b}, and it is often called a `corona'. (Note that `strongly magnetised' here refers to the fluctuating component of the magnetic field which has a pressure comparable to the thermal pressure, but this does not mean that the large-scale poloidal magnetic field is strong.) This article thus suggests that the dynamics of the corona and its connection with the transport of magnetic flux should be studied in more detail. It is unclear whether this corona should be modelled with an effective viscosity and resistivity like the main part of the disc, since some authors refer to this corona as non-turbulent due to its strong magnetisation. The vertical extension of the corona would also be of some interest for the diffusion process. Indeed, a more extended corona (because it is hot or because of magnetic support) could help reduce even more the diffusion of weak fields by increasing the length-scale on which the field lines are bent.

Finally we should stress that all this analysis relies on the assumption that the turbulence can be modelled by an effective resistivity and viscosity. It is not clear that this assumption is valid, especially for diffusive processes acting on a scale comparable to the scale-height. Indeed, in that case there might not be a real scale separation since the correlation length of the turbulence could be of the same order of magnitude. In principle then, the right method would be to perform turbulent numerical simulations that self-consistently describe the effect of turbulence. MHD simulations of a stratified disc containing a significant net vertical flux have however proved problematic so far. To avoid the further complication and the necessarily low resolution of global simulations, it would be useful to study the advection and diffusion of the magnetic flux in a local model. The classical stratified shearing box model would need some modification in order to take into account the relevant source terms. The inclusion of the non-vanishing horizontal component of the magnetic field at the surface of the disc could be achieved simply by changing the vertical boundary conditions, as has been done in this article for 1D non-turbulent simulations. A boundary condition that imposes the magnetic field may be problematic in the presence of turbulence however. The inclusion of other source terms describing the advection due to the effect of turbulence is much less obvious, but may deserve to be further investigated.

\section*{Acknowledgments}
This research was supported by STFC.  We thank Henk Spruit, Henrik Latter, Cl\'ement Baruteau, and S\'ebastien Fromang for useful discussions. We also thank S\'ebastien Fromang for providing us with the stratified shearing-box implementation of RAMSES.

\section*{Appendix A : a description of the intermediate region in the case of zero viscosity}
	\label{sec:appendix_A}
In this appendix we provide an analytical description of the intermediate region around $\zeta_B$ connecting the passive field and the force free field regimes. For this description to be tractable we assume this intermediate region to be thin, and we consider only the case of vanishing magnetic Prandtl number (i.e. zero viscosity $\alpha$, but finite resistivity $\alpha/\Pm$). We concentrate on the source terms $D_B$ and $b_{rs}$ in order to explain the bump observed in the profile of the radial component of the magnetic field (the hydrodynamical source terms with vanishing viscosity would not be very meaningful anyway). 

The density profile around $\zeta_B$ is approximated by an exponential:
\begin{equation}
\rho \simeq \frac{1}{2\beta_0}e^{-\zeta_Bx},
\end{equation}
where we defined $x\equiv\zeta-\zeta_B$. This expansion of the density profile is valid if $x\ll \zeta_B$. Note that the intermediate region where $\beta \sim 1$ has a typical thickness of $x \sim 1/\zeta_B$, therefore the assumption of a thin transition is valid if $\zeta_B^2 \gg 1$. The differential system is then rewritten in the following way:
\begin{eqnarray}
u_\phi + e^{\zeta_Bx}\left(\p_xb_r - D_B\right) & = & 0, \label{eq:app_uphi}  \\
u_r - 4 e^{\zeta_Bx}\p_xb_\phi & = & 0, \label{eq:app_ur}  \\
\p_x\left[\frac{\alpha}{\Pm}(\p_xb_r - D_B) + u_r \right]   & = & 0, \label{eq:app_br} \\
\p_x\left[\frac{\alpha}{\Pm}\p_xb_\phi + u_\phi \right]   & = & \frac{3}{2}b_r. \label{eq:app_bphi}
\end{eqnarray}
The term on the right hand side of Equation~(\ref{eq:app_bphi}) is furthermore neglected due to our assumption that the transition region is thin. Equations~(\ref{eq:app_br})-(\ref{eq:app_bphi}) then simply state that the azimuthal and radial electric field are constant, and their value can be estimated at $\zeta_Bx \rightarrow\pm\infty$ (i.e. at $\zeta=\zeta_B^\pm$): this is consistent with the boundary conditions given by equations~(\ref{eq:const_ur}) and (\ref{eq:const_uphi}). Thus
\begin{eqnarray}
u_r &=& u_\psi -\frac{\alpha}{\Pm}(\p_xb_r-D_B),  \\
u_\phi &=& \Delta u_\phi -\frac{\alpha}{\Pm}\p_xb_\phi,
\end{eqnarray}
where $u_\psi = u_r(\zeta_B^+)= \alpha/\Pm(\p_\zeta b_r(\zeta_B^-)-D_B)$ and $\Delta u_\phi = u_\phi(\zeta_B^+) = \alpha/\Pm\p_\zeta b_\phi(\zeta_B^-)$. Introducing these relations into equations~(\ref{eq:app_uphi})-(\ref{eq:app_ur}), one obtains the system of two equations:
\begin{eqnarray}
\frac{\alpha}{\Pm}\p_xb_\phi - e^{\zeta_Bx}(\p_x b_r-D_B) &=& \Delta u_\phi, \\
\frac{\alpha}{\Pm}(\p_xb_r-D_B) + 4 e^{\zeta_Bx}  \p_xb_\phi &=& u_\psi, 
\end{eqnarray}
which can be solved to obtain explicit functions for $\p_x b_r$ and $\p_x b_\phi$:
\begin{eqnarray}
\p_x b_r &=& D_B + \frac{\frac{\alpha}{\Pm}u_\psi - 4 e^{\zeta_Bx}\Delta u_\phi}{\frac{\alpha^2}{\Pm^2} + 4e^{2\zeta_Bx}}, \\
\p_x b_\phi &=& \frac{\frac{\alpha}{\Pm}\Delta u_\phi + e^{\zeta_Bx} u_\psi}{\frac{\alpha^2}{\Pm^2} + 4e^{2\zeta_Bx}}.
\end{eqnarray}
Integrating over $x$, one can compute the jump in $b_r$ across the intermediate region:
\begin{equation}
b_r(x) - b_r(-x) \simeq - \frac{\Pm\pi}{\alpha}\frac{\Delta u_\phi}{\zeta_B} + \left[\zeta_Bx - \ln\left(\frac{2\Pm}{\alpha}\right)\right]\frac{\Pm}{\alpha}\frac{u_\psi}{\zeta_B} + 2D_Bx,
\end{equation}
if $e^{\zeta_Bx} \gg 1$. The two terms proportional to $x$ represent the smooth variation in the passive field regime (from $-x$ to 0) and force free field regime (from 0 to $x$), and do not correspond to a jump in the transition region, thus 
\begin{equation} 
\Delta b_r \equiv b_r(\zeta_B^+) - b_r(\zeta_B^-) = - \frac{\Pm\pi}{\alpha}\frac{\Delta u_\phi}{\zeta_B} - \ln\left(\frac{2\Pm}{\alpha}\right)\frac{\Pm}{\alpha}\frac{u_\psi}{\zeta_B}.
\end{equation}

Similarly, one obtains the jump in $b_\phi$: 
\begin{equation}
\Delta b_\phi \equiv b_\phi(\zeta_B^+) - b_\phi(\zeta_B^-) = \frac{\Pm\pi}{4\alpha}\frac{u_\psi}{\zeta_B}  - \ln\left(\frac{2\Pm}{\alpha}\right)\frac{\Pm}{\alpha}\frac{\Delta u_\phi}{\zeta_B}
\end{equation}

These new boundary conditions can then be injected into the two-zone model. It turns out that, in the limit $\zeta_B^2\gg1$, only one term is significant, and the jump conditions can be simplified to
\begin{eqnarray}
\Delta b_r &=& - \frac{\Pm\pi}{\alpha}\frac{\Delta u_\phi}{\zeta_B},  \label{eq:jump_br} \\
\Delta b_\phi &=& 0.    \label{eq:jump_bphi} 
\end{eqnarray}
This comes from the fact that $b_r/b_\phi = O(1/\zeta_B^2)$. The two-zone model then gives the following slope of the profile of the radial component of the magnetic field:
\begin{equation}
b_{r1} = \frac{1}{1-\frac{\Pm\pi}{2\alpha}}\left(\frac{b_{rs}}{\zeta_B} + D_B \right). \label{eq:app_br1}
\end{equation}

Using the same two-zone description for the marginal antisymmetric MRI mode, one can compute the value of the magnetic diffusivity under the marginal stability hypothesis. For this purpose, one assumes the mode to be antisymmetric about the midplane (contrary to the calculation in Section~\ref{sec:anal}). We obtain
\begin{equation}
\frac{\alpha}{\Pm} = \frac{3\pi}{2},   \label{eq:alpha_marginal}
\end{equation}
which is in good agreement with the numerical calculations at large $\beta_0$ and without viscosity (respectively dashed and dotted line in Figure~\ref{fig:alpha}). Using this value of $\alpha/\Pm$ in equation~(\ref{eq:app_br1}), one finally obtains the slope of the radial magnetic field profile:
\begin{equation}
b_{r1} = \frac{3}{2}\left(\frac{b_{rs}}{\zeta_B} + D_B \right),
\end{equation}
which reproduces the larger slope observed in numerical profiles in Section~\ref{sec:diffusion}. Note that equation~(\ref{eq:app_br1}) indicates that relaxing the marginal stability hypothesis would change the slope of the radial magnetic field profile to a different factor than $3/2$. It even shows that if $\alpha/\Pm < \pi/2$ the slope and hence the transport rate would change sign ! This is not believed to be physical, and is probably an artefact of considering a stationary flow that is unstable to the MRI. This assertion is supported by the fact that the value $\alpha/\Pm=\pi/2$ corresponds to the largest-scale MRI mode \textit{symmetric} about the midplane being marginally stable (this can be obtained using the two-zone model). At still lower values of $\alpha$, the numerical profiles show multiple bending of the field lines, which is also an artefact due to the fact that the flow is unstable to the MRI \citep{ogilvie01}.

\section*{Appendix B : two-zone model with more precise boundary conditions}
	\label{sec:appendix_B}

Here we assume that the boundary conditions found in Appendix~A can be generalised to non-vanishing magnetic Prandtl number in the following way:
\begin{eqnarray}
\Delta b_r &=& - \frac{2}{3}\frac{\Delta u_\phi}{\zeta_B}, \label{eq:jump_br_heuristic} \\
\Delta b_\phi &=& 0.
\end{eqnarray}
These boundary conditions are the same as equations~(\ref{eq:jump_br}) and (\ref{eq:jump_bphi}) with the value of $\alpha/\Pm$ given by the marginal stability analysis (equation~\ref{eq:alpha_marginal}). This heuristic generalisation is motivated by the numerical results, which show that the jump in radial magnetic field is independent of the magnetic Prandtl number (see Figure~\ref{fig:cstalpha_brs_scaled_pm}). This feature is well reproduced by the above boundary conditions. Another way of looking at this generalisation is to assume a more general form inspired by equations~(\ref{eq:jump_br}) and (\ref{eq:jump_bphi}):
\begin{eqnarray}
\Delta b_r &=& - \frac{\Pm A}{\alpha}\frac{\Delta u_\phi}{\zeta_B},   \label{eq:jump_br_A}\\
\Delta b_\phi &=& 0,
\end{eqnarray}
where $A$ is some unspecified numerical factor which may depend on $\Pm$. Using then a two-zone model to describe the marginal MRI mode leads to the following value: $A=2\alpha/(3\Pm)$. Introducing this value into equation~(\ref{eq:jump_br_A}) gives equation~(\ref{eq:jump_br_heuristic}).

We now use the new boundary conditions in conjunction with the two-zone model of Section~\ref{sec:anal} to construct analytical profiles corresponding to the different source terms. The analytical profiles for the source terms $b_{r\rms}$ and $D_B$ are then
\begin{eqnarray}
b_r(\zeta<\zeta_B) &=& \frac{3}{2}\left(\frac{b_{r\rms}}{\zeta_B} + D_B \right)\zeta, \\
b_\phi(\zeta<\zeta_B) &=& \frac{3}{8}\frac{\Pm}{\alpha}\left(\frac{b_{r\rms}}{\zeta_B} +
  D_B \right)\zeta(\zeta^2-\zeta_B^2), \\
 u_{\psi} &=&  \frac{\alpha}{\Pm}\left( \frac{3}{2}
  \frac{b_{r\rms}}{\zeta_B}+  \frac{1}{2}D_B \right) ,
\label{eq:uflux_mag_bis} \\
u_\phi(\zeta>\zeta_B) &=& \frac{3}{4}b_{r\rms}\left(2|\zeta| - \zeta_B \right) +
\frac{3D_B}{4}\zeta^2.
\end{eqnarray}

Those for the source term $b_{\phi\rms}$ are
\begin{eqnarray}
b_r(\zeta<\zeta_B) &=& \frac{\alpha}{\Pm}\frac{b_{\phi\rms}}{\zeta_B^3}, \\
b_\phi(\zeta<\zeta_B) &=& \frac{b_{\phi\rms}}{4}\left(3\frac{\zeta}{\zeta_B} + \frac{\zeta^3}{\zeta_B^3}\right), \\
 u_{\psi} &=&  \frac{\alpha^2}{\Pm^2} \frac{b_{\phi\rms}}{\zeta_B^2} ,   \label{eq:uflux_bphis} \\
u_\phi(\zeta>\zeta_B) &=& \frac{3}{2}\frac{\alpha}{\Pm}\frac{b_{\phi\rms}}{\zeta_B}.
\end{eqnarray}

And finally the analytical profiles for the hydrodynamic source terms are the following:
\begin{eqnarray}
b_r &=& \frac{\Pm}{\alpha}\frac{u_{r2}}{30}\zeta(12\zeta_B^2-10\zeta^2) - \frac{2}{3}u_{\phi2}\zeta, \\
b_\phi &=& \frac{\Pm}{\alpha}(\zeta_B^2-\zeta^2)\left\lbrack\frac{2}{3}u_{\phi2}\zeta +
  \frac{u_{r2}}{120}\frac{\Pm}{\alpha}\zeta(3\zeta^2-9\zeta_B^2) \right\rbrack. \\
 u_{\psi} &=& u_{r0} + \frac{12}{30}u_{r2}\zeta_B^2 - \frac{2}{3}\frac{\alpha}{\Pm}u_{\phi2},  \label{eq:uflux_hydro} \\
u_\phi(\zeta>\zeta_B) &=& u_{\phi0} + \frac{\Pm}{\alpha}\frac{u_{r2}}{10}\zeta_B^4.
\end{eqnarray}

From these, one can deduce the inclination of the field lines at the surface of the disc, when the diffusion is compensated by advection due to effective viscosity (here for $D_H=1$, and $D_{\nu\Sigma}=0$):
\begin{equation}
\frac{B_r}{B_z} = \frac{2}{3} \frac{H}{r}\Pm\zeta_B\left\lbrack  -\frac{1}{2}-\frac{1}{6\Pm} +  \frac{2}{4\alpha^2+1}\left((4-\frac{2}{3\Pm})\alpha^2+\frac{12\zeta_B^2}{30} \right) \right\rbrack,
 \end{equation}
as well as the ratio of the radial to azimuthal components of the surface magnetic field, when the diffusion is compensated by the transport due to an outflow:
\begin{equation}
\frac{B_{r\rms}}{B_{\phi\rms}} = - \frac{u_{b\phi\rms}}{u_{br\rms}} = -\frac{2}{3}\frac{\alpha}{\Pm}\frac{1}{\zeta_B^2}.
	\label{eq:inc_bphis}
\end{equation}

\bibliography{discs}

\bsp
\label{lastpage}

\end{document}